\documentclass[12pt]{iopart}
\usepackage{iopams,ulem,color,graphicx,url}
\begin{document}
\def\be{\begin{equation}}
\def\ee{\end{equation}}
\def\bea{\begin{eqnarray}}
\def\eea{\end{eqnarray}}
\def\erf{\mbox{\rm Erf}}
\def\bea{\begin{eqnarray}}
\def\eea{\end{eqnarray}}
\title{Comparison of two models of tethered motion}
\author{Luca Giuggioli$^{1,2}$, Shamik Gupta$^3$, Matt Chase$^4$}
\address{$^1$\mbox{Bristol Centre for Complexity Sciences, University of Bristol, Bristol,
UK}\\
$^2$Department of Engineering Mathematics, University of Bristol, Bristol,
UK\\
$^3$Department of Physics, Ramakrishna Mission Vivekananda University, Belur Math, Howrah 711202, India\\
$^4$Department of Physics and Astronomy, University of New Mexico, Albuquerque, NM, USA}
\ead{Luca.Giuggioli@bristol.ac.uk}
\vspace{10pt}
\begin{abstract}
We consider a random walker whose motion is tethered around a focal
point. We use two models that exhibit the same spatial dependence in the
steady state but widely different dynamics. In one case, the walker is
subject to a deterministic bias towards the focal point, while in the
other case, it resets its position to the focal point at random times.
The deterministic tendency of the biased walker makes the forays away
from the focal point more unlikely when compared to the random nature of
the returns of the resetting walker. This difference has consequences on
the spatio-temporal dynamics at intermediate times. To show the
differences in the two models, we analyze their probability distribution
and their dynamics in presence and absence of partially or fully
absorbing traps. We derive analytically various quantities: (i) mean
first-passage times to one target, where we recover results obtained
earlier by a different technique, (ii) splitting probabilities to either of two targets as well as survival probabilities when one or either target is partially absorbing. The interplay between confinement, diffusion and absorbing traps produces interesting non-monotonic effects in various quantities, all potentially accessible in experiments. The formalism developed here may have a diverse range of applications, from study of animals roaming within home ranges and of electronic excitations moving in organic crystals to developing efficient search algorithms for locating targets in a crowded environment.
\end{abstract}
\vspace{2pc}
\noindent{\it Keywords}: 

\section{Introduction}
Tethered motion has relevance in a variety of context and represents the
inability of a particle or agent to stray afar from a given location in
space, the so-called focal point. Animals that keep returning to
their burrow for shelter or for caching resources, or animals who do
not have a den but who move away from predators or other conspecifics and
tend to return to the same area are ecological examples
\cite{giuggiolikenkre2014}. Other worthwhile examples to cite are
searchers that reset their position to their initial starting place
after being unsuccessful for some time \cite{kusmierzetal2014}, or a
cell population that suffers an abrupt event and its density gets
reduced to some specified lower value, e.g., tumors after chemical treatment \cite{viscoetal2010}.

Although in all the above cases, the steady-state probability is
expected to be localized around a focal point or value, the difference
between the first and the last two examples cited above is that an agent moves towards the focal point by going through the intervening space in the first case, whereas it does not do so in the last two cases. 
This difference has two important consequences. Firstly, the formalism
for analytically characterizing the former is a Smoluchowski-type
diffusion equation, while that for the latter is the diffusion equation
with stochastic resetting \cite{evansmajumdar2011a}. Secondly, the
steady-state probability distribution for the animal example is an equilibrium state, while it is not so for the resetting case where detailed balance does not hold \cite{evansmajumdar2014}.

Analysis of the resetting example has shown that the steady-state
probability distribution is an exponential symmetric around the focal
point \cite{evansmajumdar2011a,evansmajumdarmallick2013}. As the spatial dependence in the steady
state of the Smoluchowski-type diffusion equation depends on the
deterministic tendency towards the focal point, we choose for comparison
a model where the steady state is also a symmetric exponential, which
corresponds to a motion subject to a constant bias towards the focal point.

As the spatial distribution at long times for the two models coincides, the fundamental difference of the two processes should clearly appear in the dynamics. Here we are interested in unraveling such differences by comparing the models in one dimension (1D). For comparison, we discuss the form of the propagators and some of their moments, and also discuss applications to two reaction-diffusion scenarios.

The paper is laid out as follows. In Sec. \ref{sec:models},
we describe the two tethered models that we will use for comparison. We
name drift model the case where the motion is subject to a constant deterministic bias towards the focal point, to distinguish it from the
(sudden) resetting model. We will present exact analytical expressions for
the spatial distribution, and analyze in detail the time dependence of
the mean and the mean-squared displacement from the focal point. Section
\ref{sec:traps} is dedicated to the formalism to treat the case in which
the motion takes place in presence of a trap. We also show how to
compute the yield, that is, the number of particles getting absorbed at
a detector placed at the trap location, when the motion represents the movement of a particle decaying over time.
In
Sec. \ref{sec:msd_mfpt}, we study in all three cases the parameter
dependence of the mean first-passage time to a given target. In Sec.
\ref{sec:split}, we consider the scenario of two targets, construct the probability distribution and
investigate aspects related to the splitting probability, namely, the
probability of reaching one of the targets (without having been at the
other) as a function of the initial distance from it. Finally, Sec. \ref{sec:concl} constitutes the concluding section.

\section{Focal point models}
\label{sec:models}
The common characteristic of the two models is that
the uncertainty in the movement of the stochastic observable through space is represented by
Brownian diffusion. We may therefore refer to the stochastic variable as
representing the state of a random walker. The tendency or the bias to move towards the focal point on the other
hand distinguishes the two models qualitatively. In the case of the
drift model, the bias is represented by a constant drift towards the
focal point that is present {\it at all times}, while in the case of the stochastic resetting model, the bias is
represented through a long jump (an infinitely fast movement) to the
focal point {\it at random times}. The former is an example of a
Smoluchowski-type model that has been used in various contexts, e.g., to model Brownian particles subject to dry friction
\cite{touchetteetal2010} and motion of excitons in doped molecular crystals and signal receptor clusters on the surface of T-cells during immunological synapse formation \cite{chaseetal2016}. It has also appeared in the animal movement literature to represent movement within a home range, and is called the Holgate-Okubo model \cite{holgate1971,okubobook1980,okubogross2001}. The stochastic resetting model was introduced more recently to model search processes in a crowded environment \cite{evansmajumdar2011a}. 

In the following, we consider the motion of the walker to be taking place in one dimension. We take $(x,t)$ to represent space and time, respectively, and consider the constant $D>0$ to represent the diffusion constant, $x_0$ to represent the initial location of the walker and $x_c$ to denote the focal point to which the motion of the walker is biased. In the constant drift model, the position of the walker evolves in time according to the Langevin equation of motion
\be
\frac{{\rm d}x}{{\rm d}t}=-\frac{\partial V(x)}{\partial x}+\sqrt{2D}~\eta(t),
\ee
where $\eta(t)$ is a Gaussian, white noise with zero mean and with correlation given by
\be
\langle \eta(t)\eta(t')\rangle=\delta(t-t').
\ee
Here and in the following, angular brackets denote averaging over noise realizations. We take the potential $V(x)$ to be of the form $V(x)=v|x-x_c|$, where the parameter $v$ represents a speed. In the stochastic resetting model, while at position $x$ at time $t$, the walker in the ensuing infinitesimal time interval ${\rm d}t$ has the following choices for updating its location: 
\bea 
&&\mathrm{With~probability~}1-rdt, \mathrm{~ following~the~dynamics~} \frac{{\rm d}x}{{\rm d}t}=\sqrt{2D}~\eta(t), \\
&&\mathrm{With~probability~}rdt, \mathrm{~resetting~to~the~focal~point~}x_c.
\eea
Here, $r\ge 0$ is a parameter that characterizes the rate of resetting, that is, the probability to reset per unit time.  The
dynamics of the resetting model involves indeed a restart or a ``reset" of the
motion from $x_c$ at random times, where $x_c$ may be viewed as the reset
location.  From the foregoing, it follows that the time interval $\tau$ between successive resets is a random variable distributed according to an exponential distribution $\rho(\tau)=re^{-r\tau}$.  

\begin{figure}[!h]
\includegraphics[width=150mm]{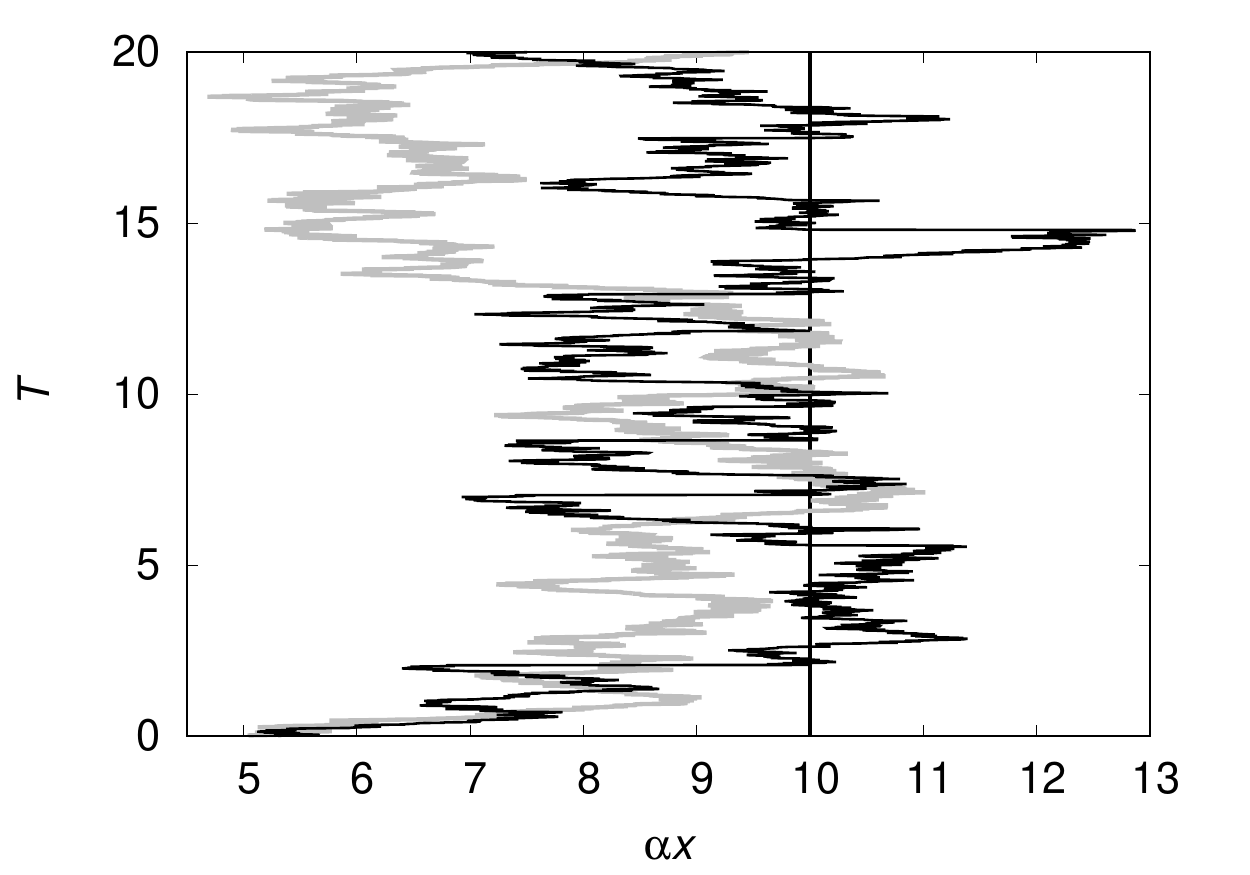}
\caption{Representative trajectories for the constant drift model and
the stochastic resetting model. The black trajectory refers to stochastic resetting model, while the grey one corresponds to the drift model. The values of the dynamical parameters
of the two models are chosen so as to have the steady-state
characteristic spatial scale in the two models the same, by setting
$v/D=\sqrt{r/D}=\alpha$, see Eqs. (\ref{eqn:ss-V}) and
(\ref{eqn:steadystate-exp}). The temporal scale is on the other hand set
to $rt=T$ for the drift model and $v^2t/D=T$ for the resetting model.
Here, we take $\alpha x_0=5$ and $\alpha x_c=10$, with the latter
indicated in the plot with a vertical line. Notice the presence of horizontal segments in the black trajectory corresponding to time instants when the walker resets to $x_c$, which are instead lacking in the grey trajectory for the constant drift model.}
\label{fig:reset-schematic}
\end{figure}

To provide a demonstration of the difference between the two models, we
plot in Fig.~\ref{fig:reset-schematic} two typical trajectories, one for the constant drift model and one for the resetting model. 

\subsection{The drift model}
The probability distribution $P(x,t|x_0,0)$ to find the walker at position $x$ at time $t$, given that it was at $x_0$ at time $t=0$, that is $P(x,0)=\delta(x-x_0)$, where $\delta(z)$ represents a Dirac delta, is governed by the Fokker-Planck (FP) equation
\be
\frac{\partial P(x,t)}{\partial t}=v\frac{\partial }{\partial
x}\left[\frac{x-x_c}{|x-x_c|}P(x,t)\right]+D\frac{\partial^2P(x,t)}{\partial x^2}.
\label{eqn:FP-HO}
\ee
Exact solution of Eq.~(\ref{eqn:FP-HO}) for any $x$ and $x_0$ can be
found in two ways. A separation of variable allows to map the
problem to an eigenvalue problem of the Sturm-Liouville type
\cite{touchetteetal2010}. Another way is to solve the equation in the
Laplace variable $\epsilon$, that is, to find $\widetilde{P}(x,\epsilon|x_0,0)\equiv \int_0^\infty {\rm d}t~
e^{-\epsilon t}P(x,t|x_0,0)$ in the region to the left and to the right
of $x_0$, and then match the two solutions at $x_0$ to obtain \cite{chaseetal2016}
\be
\fl \widetilde{P}(x,\epsilon|x_0,0)=\frac{e^{-\frac{v(|x-x_c|-|x_0-x_c|)}{2D}}}{v\sqrt{1+\frac{4D\epsilon}{v^2}}}\left[e^{-\frac{v|x-x_0|\sqrt{1+4D\epsilon/v^2}}{2D}} +\frac{e^{-\frac{v\left(|x-x_c|+|x_0-x_c|\right)\sqrt{1+4D\epsilon/v^2}}{2D}}}{\sqrt{1+\frac{4D\epsilon}{v^2}}-1}
\right].
\label{eqn:Lap-HO}
\ee
Inverse Laplace transform of the above expression then yields \cite{chaseetal2016}
\bea
&& P(x,t|x_0,0)=\frac{1}{\sqrt{4\pi D
t}}e^{-\frac{(x-x_0)^2}{4Dt}}e^{-\frac{v^2t}{4D}}e^{-\frac{v(|x-x_c|-|x_0-x_c|)}{2D}}
\nonumber \\&&+\frac{v e^{-\frac{v |x-x_c|}{D}}}{4D}\left[1-\erf\left(\frac{|x-x_c|+|x_0-x_c|-v
t} {\sqrt{4Dt}}\right)\right],
\label{eqn:time-HO}
\eea
with $\erf(x)=(2\pi)^{-1}\int_{0}^{x}ds\,e^{-s^2}$ being the error
function. In the limit $t\rightarrow +\infty$, Eq.
(\ref{eqn:time-HO}) yields the steady state distribution 
\be
P_{\rm ss}(x)=\frac{v}{2D}e^{-v|x-x_c|/D}.
\label{eqn:ss-V}
\ee

From Eq. (\ref{eqn:time-HO}), we obtain the mean position as
\bea
&&\fl \langle x-x_0\rangle(t)=x_c-x_0+\frac{vt-(x_c-x_0)}{2}{\rm Erfc}\left[\frac{1}{2}\left(\sqrt{\frac{t}{D}}v-\frac{x_c-x_0}{\sqrt{Dt}}\right)\right]e^{\frac{v\left[|x_c-x_0|-(x_c-x_0)\right]}{2D}} \nonumber \\
&&\fl -\frac{vt+(x_c-x_0)}{2}{\rm Erfc}\left[\frac{1}{2}\left(\sqrt{\frac{t}{D}}v+\frac{x_c-x_0}{\sqrt{Dt}}\right)\right]e^{\frac{v\left[|x_c-x_0|+(x_c-x_0)\right]}{2D}},
\label{eqn:mean-HO}
\eea
where ${\rm Erfc}(x)\equiv 1-{\rm Erf}(x)$ is the complementary error
function. Equation (\ref{eqn:mean-HO}) is identically zero for
$x_0=x_c$, is non-zero for any finite $t$ if $x_0\not= x_c$, and equals
$x_c-x_0$ at long times. The calculation of the mean-squared
displacement (MSD) is more involved, but can be done analytically. While
we have reported the full general expression in Appendix A, for the
simple case of $x_0=x_c$, we have
\be
\fl \langle (x-x_c)^2\rangle(t)=2\frac{D^2}{v^2}-\left[\sqrt{\frac{Dt}{\pi}}\left(vt+2\frac{D}{v}\right)\right]e^{-\frac{v^2t}{4D}}+\left(2Dt+\frac{v^2t^2}{2}-2\frac{D^2}{v^2}\right){\rm Erfc}\left(\frac{1}{2}\sqrt{\frac{t}{D}}v\right).
\label{eqn:msd_centred}
\ee
While in the general case, i.e., for $x_0\not=x_c$, the short time
dependence of the MSD is $2Dt+v^2t^2$, which is the expected time
dependence of a diffusing particle subject to a drift, this is not the
case for $x_0=x_c$. At short times, in fact, Eq. (\ref{eqn:msd_centred})
increases proportionally to
$2Dt-\frac{8}{3}\sqrt{\frac{v^2Dt^3}{\pi}}+\frac{v^2t^2}{2}$. This is
because when a walker starts at the centre, the potential acts to
confine the walker around the origin rather than forcing it towards
$x_c$. As the walker diffuses outward, the potential should
reduce the MSD depending on its strength $v$ and the rate at
which the walker diffuses that depends on $D$. This is why the expected drift $v^2t^2$ is reduced by a factor two and the additional contribution proportional to $t^{3/2}$ is negative.

\subsection{The stochastic resetting model}
We now discuss the model of stochastic resetting. Recognizing that at each reset, the motion starts afresh
(gets ``renewed") at $x_c$, we may straightforwardly obtain the probability
$P(x,t|x_0,0)$. At a fixed time t, let the time elapsed since the last renewal be in
$[\tau, \tau +{\rm d}\tau]$, with $0 \le \tau \le t$. Noting that the probability for
this event is $r\exp(-r\tau){\rm d}\tau$, we have
\be
\fl P(x,t|x_0,0)=\int_0^{t}{\rm d}\tau~r\exp(-r\tau)P_{\rm
d}(x,t|x_c,t-\tau)+\exp(-rt)P_{\rm d}(x,t|x_0,0),
\label{eqn:pxt-gen}
\ee
where 
$P_{\rm d}(x,t|x_c,t-\tau)\equiv e^{-(x-x_c)^{2}/(4D\tau)}/\sqrt{4\pi
D\tau}$ is the propagator of free
diffusion.
Integrating Eq. (\ref{eqn:pxt-gen}) over $x \in
[-\infty,\infty]$, and using the normalization $\int_{-\infty}^\infty
{\rm d}x~P_{\rm d}(x,t|x_0,t')=1~\forall~x_0$ and for $t>t'$, it may be checked that $P(x,t|x_0,0)$ is correctly normalized to unity.
\begin{figure}[!ht]
\includegraphics[width=150mm]{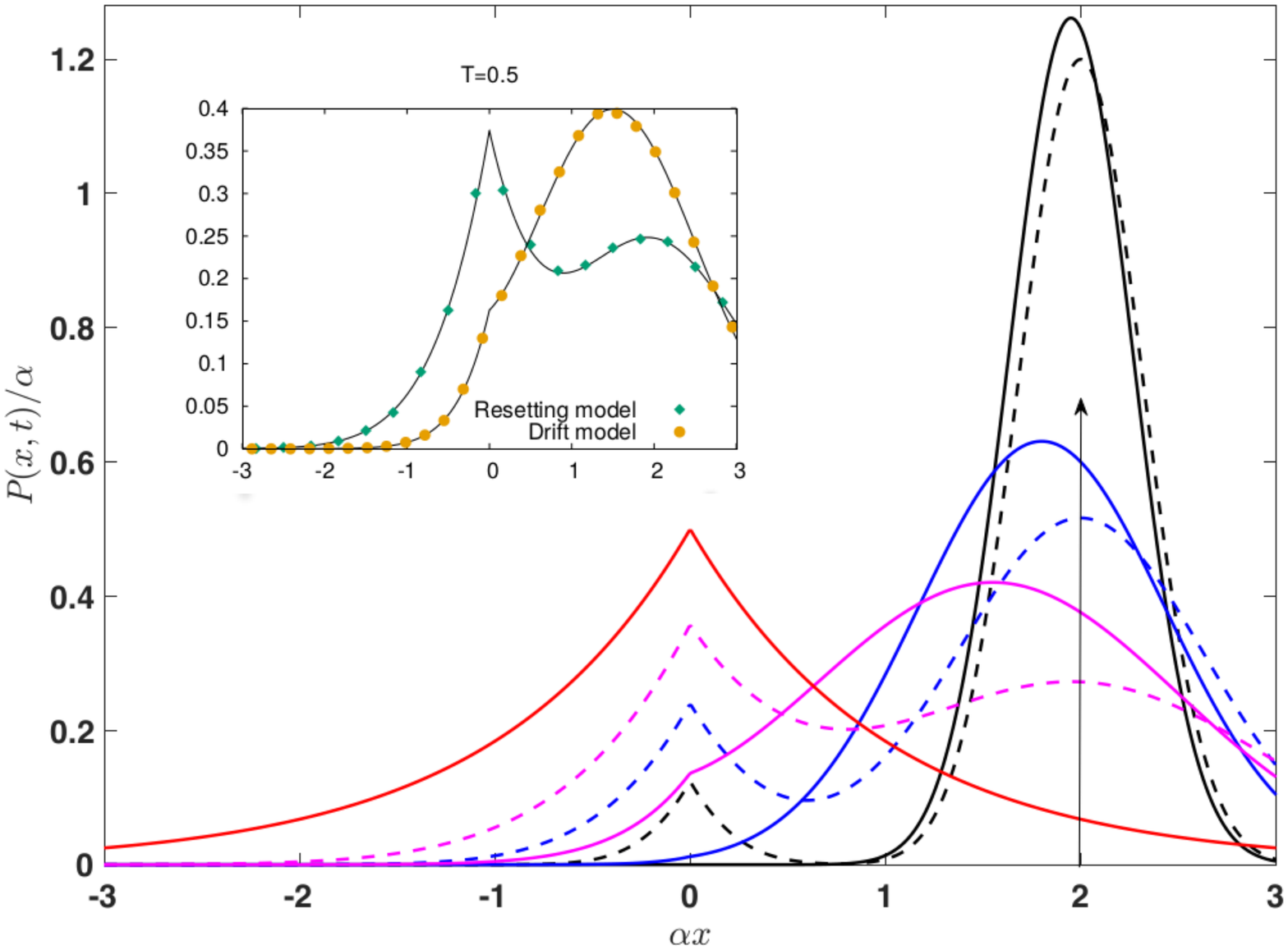}
\caption{(Color online) Comparison of the time-dependent probability
distribution for the drift model (solid line) in Eq.
(\ref{eqn:time-HO}), and the resetting model (dashed line) in Eq.
(\ref{eqn:prop_reset_exp}). The arrow corresponds to the localized
initial position at $\alpha x_0=2$, while the focal point is set at
$\alpha x_c=0$. The solid curve symmetric around the origin is the
steady state distribution, identical for both models, while the other
curves, starting from the highest curves on the right, represent the two models at times $T=$0.05, 0.45, and 9.95. The
        inset shows the match between theory and simulations for
        $T=0.5$. The parameter $\alpha$ and the rescaled time $T$ are
        the same as in Fig. \ref{fig:reset-schematic}. }
\label{fig:propagators}
\end{figure}
From Eq. (\ref{eqn:pxt-gen}), we obtain 
\bea
&& P_{\mbox{reset}}(x,t|x_0,0)=\frac{e^{-rt-\frac{(x-x_0)^2}{4Dt}}}{\sqrt{4\pi Dt}}+\frac{1}{4}\sqrt{\frac{r}{D}}e^{-\sqrt{\frac{r}{D}}|x-x_c|}{\rm Erfc}\left(\frac{|x-x_c|}{2\sqrt{Dt}}-\sqrt{rt}\right) \nonumber \\
&& -\frac{1}{4}\sqrt{\frac{r}{D}}e^{\sqrt{\frac{r}{D}}|x-x_c|}{\rm Erfc}\left(\frac{|x-x_c|}{2\sqrt{D
t}}+\sqrt{rt}\right),
\label{eqn:prop_reset_exp}
\eea
whose steady-state form is
\be
P_{\rm ss}(x)=\frac{1}{2}\sqrt{\frac{r}{D}}e^{-|x-x_c|\sqrt{r/D}}.
\label{eqn:steadystate-exp}
\ee
As expected, the steady-state distribution does not depend on the initial location $x_0$. 
For the particular case when the initial location $x_0$ is the same as
the reset location $x_c$, Eq. (\ref{eqn:steadystate-exp}) reduces to the
result obtained in Ref. \cite{evansmajumdar2011a}.
\begin{figure}[h]
\includegraphics[width=150mm]{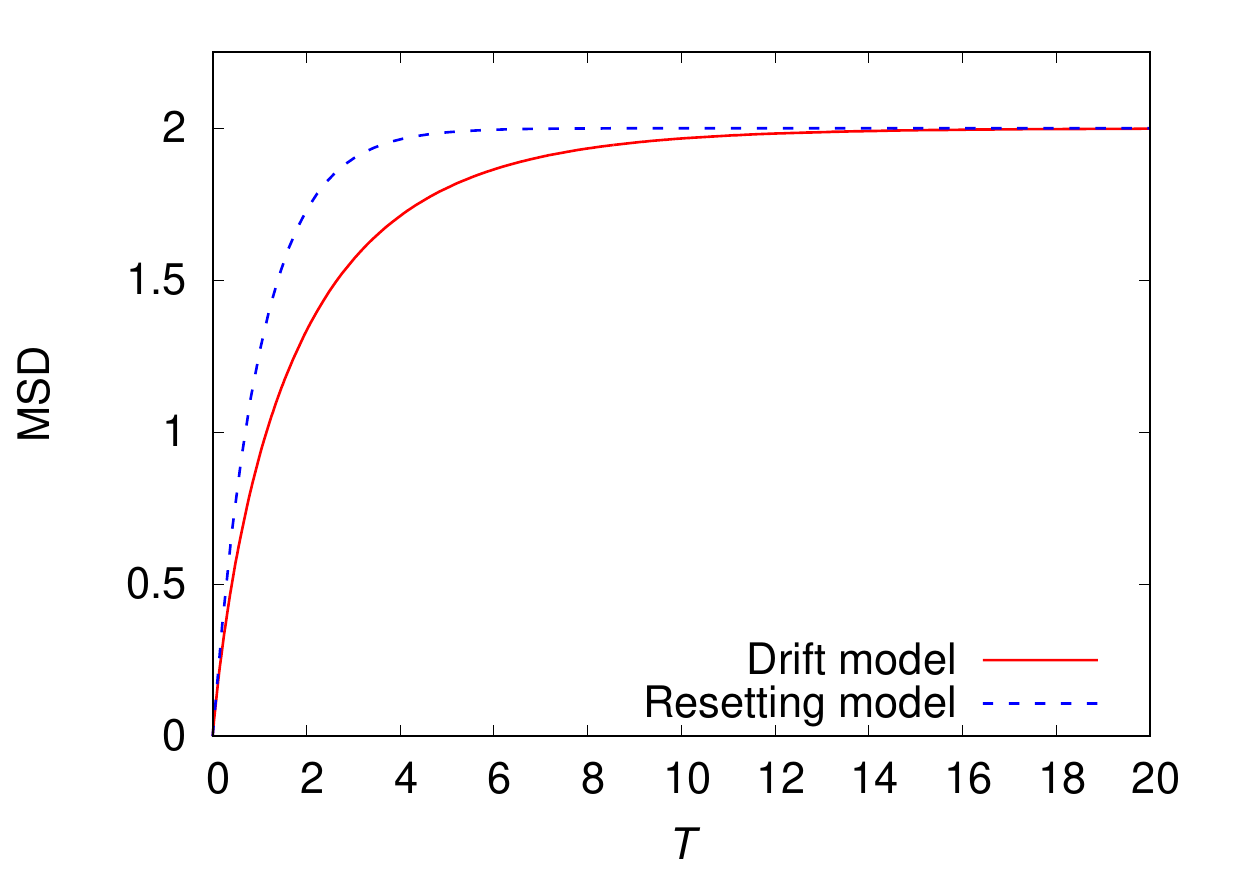}
\caption{(Color online) MSD in the drift model and the resetting model, given respectively by Eqs. (\ref{eqn:msd_centred}) and  (\ref{eqn:msd_reset_exp}) with $x_c=x_0$, as a function of the dimensionless time $T$. We used $\alpha=1$ for both models.}
\label{fig:msd}
\end{figure}

Corresponding to the distribution (\ref{eqn:prop_reset_exp}), the mean displacement from the initial location is given by
\be
\langle (x-x_0)\rangle(t)=\langle(x-x_0)\rangle'(t),
\label{eqn:mean_exp_reset}
\ee
where the prime denotes averaging with respect to the second and the
third term on the right hand side (rhs) of Eq.~(\ref{eqn:prop_reset_exp}),  and we have used
the fact that when averaged with respect to the first term, the mean displacement from the initial location evaluates to zero.  Next, we rewrite the rhs of Eq. (\ref{eqn:mean_exp_reset}) to obtain 
\be
\langle(x-x_0)\rangle(t)=
\langle (x-x_c)\rangle'(t)+
(x_c-x_0)\langle 1 \rangle'(t)=(x_c-x_0)(1-e^{-rt}),
\ee
where in obtaining the last equality, we have used the fact that
$\langle (x-x_c)\rangle'(t)=0$ because of the symmetry of the second and
the third term on the rhs of Eq. (\ref{eqn:prop_reset_exp}) under $x\to -x, x_c \to -x_c$, and that $\langle 1 \rangle'=1-e^{-rt}$ 

Similarly, one may compute the MSD as a function of time: 
\be
\fl \langle (x-x_0)^2\rangle(t)=\langle (x-x_c)^2\rangle'(t)+
(x_c-x_0)^2\langle 1 \rangle'(t)+2(x_c-x_0)\langle (x-x_c)\rangle'(t)+2Dte^{-rt},
\ee
where the fourth term on the
right is the result obtained by averaging with respect to the first
term on the rhs of Eq. (\ref{eqn:prop_reset_exp}). We finally get 
\bea
\langle (x-x_0)^2\rangle(t)
&=&\frac{2D}{r}\left[1-e^{-rt}(1+rt)\right]+(x_c-x_0)^2(1-e^{-rt})+2Dte^{-rt}\nonumber \\
&=&\left[\frac{2 D}{r}+(x_c-x_0)^2\right]\left(1-e^{-rt}\right).
\label{eqn:msd_reset_exp}
\eea

To find the $\epsilon$ expression for the propagator, it is straightforward to Laplace transform Eq. (\ref{eqn:pxt-gen}), rather than Eq. (\ref{eqn:prop_reset_exp}), to obtain
\be
\widetilde{P}(x,\epsilon|x_0,0)=\frac{e^{-|x-x_0|\sqrt{(r+\epsilon)/D}}}{2\sqrt{D(r+\epsilon)}}+\frac{r}{\epsilon}\frac{e^{-|x-x_c|\sqrt{(r+\epsilon)/D}}}{2\sqrt{D(r+\epsilon)}}.
\label{eqn:lap_reset_exp}
\ee

For convenience that will become evident in Sec. \ref{sec:split}, it is better to have the following compact expression for the Laplace propagators:
\be
\widetilde{P}(x,\epsilon|x_0,0)=h_{x_c}(x,x_0)\frac{\mathcal{C}(\epsilon)e^{-\alpha|x-x_0|\mathcal{S}(\epsilon)}+\mathcal{D}_{x_c}(x_0,\epsilon)e^{-\alpha|x-x_c|\mathcal{S}(\epsilon)}}{2\mathcal{C}(\epsilon)\mathcal{S}(\epsilon)},
\label{eqn:reset_HO_lap}
\ee
with
\bea
\fl &&\alpha=\frac{v}{D},\qquad h_{x_c}(y,z)=\frac{e^{-\alpha
(|y-x_c|-|z-x_c|)/2}}{v},\qquad
\mathcal{D}_{x_c}(y,\epsilon)=e^{-\alpha|y-x_c|\mathcal{S}(\epsilon)},\nonumber
\\ && \mathcal{S}(\epsilon)=\frac{\sqrt{1+\frac{4D\epsilon}{v^2}}}{2}, \qquad\mathcal{C}(\epsilon)=2\mathcal{S}(\epsilon)-1,
\label{eqn:cond_HO_lap}
\eea
for the drift model, and with
\bea
\fl \alpha=\sqrt{\frac{r}{D}}, \qquad
h_{x_c}(y,z)=\frac{1}{\sqrt{Dr}},\qquad
\mathcal{D}_{x_c}(y,\epsilon)=1,\qquad
\mathcal{S}(\epsilon)=\sqrt{1+\frac{\epsilon}{r}},\qquad\mathcal{C}(\epsilon)=\frac{\epsilon}{r},
\nonumber \\
\label{eqn:cond_reset_lap}
\eea
for the resetting model.

A comparison of the dynamics of the probability distribution of the two
models in Fig. \ref{fig:propagators} shows how in the 
resetting model the random walker moves through the
intervening space so that the probability distribution at short times
remains close to zero in the space between the origin and the initial
position $x_0$. From Fig. \ref{fig:propagators}, it is also evident that
far from the initial position $x_0$, the probability distribution for the
drift model is always smaller than the corresponding one for the reset
model. The limited dispersion of the drifting versus the resetting
walker is more clearly visible when the initial location is $x_0=x_c$.
In such a case, it is sufficient to plot the MSD and notice that its
value for the reset model is always larger than that for the drift model (see Fig. \ref{fig:msd}).

\section{Traps, impartial detection and yield}
\label{sec:traps}
In many scenarios, the walker dynamics may be subject to physical
constraints. When certain locations in the space the walker is roaming
about hinder
the movement, the landscape may be represented by partially absorbing
traps that capture the walker for some time \cite{kenkreetal2009}. The
walker may also represent a searcher seeking a target that is partially
hidden, and so the capture of that target may not occur upon reaching
the target for the first time but only after a larger number of
encounters. In this case, one is interested in estimating the
probability of capture as a function of time. In both these cases, the
walker may undergo, in addition, a decay before
the absorption at the trap is complete or before the target is detected,
e.g., a diffusing molecular excitation with a radiative decay in presence of a trap with imperfect capture \cite{wolf1968} or an animal being predated while foraging \cite{limadill1990}. To help interpret each parameter of the system when one is using either of the two models in the context of a forager/searcher or an electronic excitation, we have created Table \ref{tab:explan}.
\begin{table}[h]
\label{tab:explan}
\begin{center}
\resizebox{\textwidth}{!}{
\begin{tabular}{ | l | c | c | c | c |}
\hline
 & $C$ & $v$ & $r$ & $\tau$ \\
 \hline
 electronic & absorption rate & strength attracting & reset rate to & excitation \\
 excitation & of detector  & centre & attracting centre & decay rate \\
 \hline
foraging & detection rate of & movement bias to & jump return rate & predation   \\
  animal & concealed food items & home range centre & to home range centre & rate \\
 \hline
\end{tabular}}
\caption{Interpretation of the different model parameters depending on system being modelled in Eq. (\ref{eqn:FP-HO_traps}) and (\ref{eqn:FP_reset_traps}).}
\end{center}
\end{table}

To account for the aforementioned dynamical scenarios, it is necessary
to augment the equation for the spatial probability distribution with
two additional terms: a probability loss with a decay rate $\tau^{-1}$
and a rate of absorption $C$ (per unit distance) at the traps, or, a
rate $C$ of imperfect detection of the target. For the simple case of one
trap or target located at $x'$, we have
\be
\fl \frac{\partial \mathcal{P}(x,t)}{\partial
t}+\frac{\mathcal{P}(x,t)}{\tau}=v\frac{\partial }{\partial
x}\left[\frac{x-x_c}{|x-x_c|}\mathcal{P}(x,t)\right]+D\frac{\partial^2\mathcal{P}(x,t)}{\partial
x^2}-C\delta(x-x')\mathcal{P}(x,t),
\label{eqn:FP-HO_traps}
\ee
with the initial condition $P(x,0)=\delta(x-x_0)$ for the constant drift model, and
\be
\fl \mathcal{P}(x,t|x_0,0)=e^{-\frac{t}{\tau}}P(x,t|x_0,0)-C\int_{0}^{t}ds\,
e^{-\frac{t-s}{\tau}}P(x,t-s|x',0)\mathcal{P}(x',s|x_0,0)
\label{eqn:FP_reset_traps}
\ee
for the resetting model. While the former equation follows the standard
literature \cite{kenkre1982}, the latter is being reported here for the
first time. In Eq. (\ref{eqn:FP_reset_traps})  the first term contains
the defect-free propagator $P(x,t|x_0,0)$, given by Eq.
(\ref{eqn:prop_reset_exp}), multiplied by $\exp(-t/\tau)$ to represent
the loss of probability irrespective of the spatial position. The second
term indicates instead that there is a loss of probability with rate $C$
if the walker was at $x'$ at some earlier time $s$, which is a term
proportional to the probability $\mathcal{P}(x',s|x_0,0)$. As the equation
describes the probability of being at $x$ as a function of time $t$,
$\mathcal{P}(x',s|x_0,0)$ is multiplied with the probability of moving from
$x'$ to $x$ in the interval $t-s$, that is, with
$e^{-(t-s)/\tau}P(x,t-s|x',0)$. As the probability loss due to the
walker being at $x'$ may occur at any earlier time $s$, the term
$e^{-(t-s)/\tau}P(x,t-s|x',0)\mathcal{P}(x',s|x_0,0)$ is then integrated over all times $s$.

Given the linearity of Eqs. (\ref{eqn:FP-HO_traps}) and
(\ref{eqn:FP_reset_traps}) the exact solution in Laplace domain can be
obtained by application of the so-called defect technique
\cite{montrollpotts1955,kenkre1980}, which allows to write the
propagator $\mathcal{P}(x,t|x_0,0)$ of the defect problem in terms of
the defect-free propagator $P(x,t|x_0,0)$. By exploiting the linearity in probability space and realizing that the second term in Eqs. (\ref{eqn:FP-HO_traps}) and the exponential $\exp(-t/\tau)$ in Eq.  (\ref{eqn:FP_reset_traps}) simply transform the functional form of the propagator (with defect) from being expressed in terms of $\epsilon$ to being expressed in terms of $\epsilon +1/\tau$, one has for both models
\be
\fl \widetilde{\mathcal{P}}(x,\epsilon|x_0,0)=\widetilde{P}\left(x,\epsilon+\tau^{-1}|x_0,0\right)-\frac{\widetilde{P}\left(x,\epsilon+\tau^{-1}|x',0\right)\widetilde{P}\left(x',\epsilon+\tau^{-1}|x_0,0\right)}{\frac{1}{C}+\widetilde{P}\left(x',\epsilon+\tau^{-1}|x',0\right)}.
\label{eqn:prop_defects}
\ee
The terms on the rhs depend in both cases on the defect free propagator,
which are given explicitly by Eqs. (\ref{eqn:reset_HO_lap}) and
(\ref{eqn:cond_HO_lap}) for the drift model, and by Eqs. (\ref{eqn:reset_HO_lap}) and (\ref{eqn:cond_reset_lap}) for the resetting model.

We proceed by adopting quantum yield calculations \cite{wolf1976,
smithetal1979} that were used in the understanding of the motion of
Frenkel excitons in doped molecular crystals \cite{wolf1967,
powell1975}.  Electronic excitations, generated by light impulses, move
through the material and are subject to either radiative decay or
absorption by one of a small number of guest molecules introduced into
the host crystal \cite{fortetal1983}. Though quantum mechanical in
nature, the excitations were found to move incoherently on the lattice
\cite{kenkre1982} due to the weak coupling between the organic crystals
\cite{wolf1967, powell1975}.  The resulting equations that describe
their dynamics are therefore for \textit{probabilities}, and not for amplitudes, and are in the form given by
	\be
	\frac{\partial \mathcal{P}_e(x,t)}{\partial t}+\frac{\mathcal{P}_e(x,t)}{\tau_e}=\left\lbrace\mbox{Motion Terms}\right\rbrace-C_e\delta(x-x_e)\mathcal{P}_e(x,t),
	\label{eqn:exciton}
	\ee
where $\left\lbrace\mbox{Motion Terms}\right\rbrace$ describes the motion of the excitation in the absence of the guest molecules and is generally diffusive \cite{wolf1976, kenkre1982}.  Effects of the radiative decay and capture are specified by the parameters $\tau_e$, $C_e$, and $x_e$, which, respectively, represent the excitation decay time, the capture rate (per unit distance) and the location of the guest site.

The quantum yield provides a useful timescale with which to compare
empirical findings.  It is found by integrating $\mathcal{P}_e(x,t)$
over all space that
\be
\widetilde{\phi}(\epsilon)=\int_{-\infty}^{\infty}dx \widetilde{\mathcal{P}}_e(x,\epsilon)
\label{eqn:yield_0}
\ee
represents the ratio of the number of excitations that have been
absorbed by a guest molecule and radiated from the host to those that
were originally created in the host. Although the quantum yield is in
general time dependent, its value integrated over all times, that is, $\widetilde{\phi}(0)$, provides an effective rate that can be used in more complex models.

Though we do not propose that our formalism is necessarily applicable to
the study of motion of excitations in doped molecular crystals, the
quantum yield style calculations outlined above have found recent use in
other systems. For example, in the context of Smoluchowski random
walkers \cite{spendieretal2013} and in the transmission of infectious
epidemics \cite{kenkresugaya2014}, it was demonstrated that the interplay
between diffusion, confinement and absorption at site $x'$ may bring
about non-monotonic dependencies of the quantum yield.  We explore below such non-monotonic dependencies in our formalism.

With the addition of localized absorption and decay, Eq. (\ref{eqn:FP-HO_traps}) is transformed to
\bea
&&\frac{\partial \mathcal{P}_e(x,t)}{\partial
t}+\frac{\mathcal{P}_e(x,t)}{\tau}+\frac{\mathcal{P}_e(x,t)}{\tau_e} = v\frac{\partial }{\partial
	x}\left[\frac{x-x_c}{|x-x_c|}\mathcal{P}_e(x,t)\right]\nonumber\\&&+D\frac{\partial^2\mathcal{P}_e(x,t)}{\partial x^2}-C\delta(x-x')\mathcal{P}_e(x,t)-C_e\delta(x-x_e)\mathcal{P}_e(x,t).
\label{eqn:FP-HO_exciton}
\eea
\begin{figure}[h]
	\includegraphics[width=150mm]{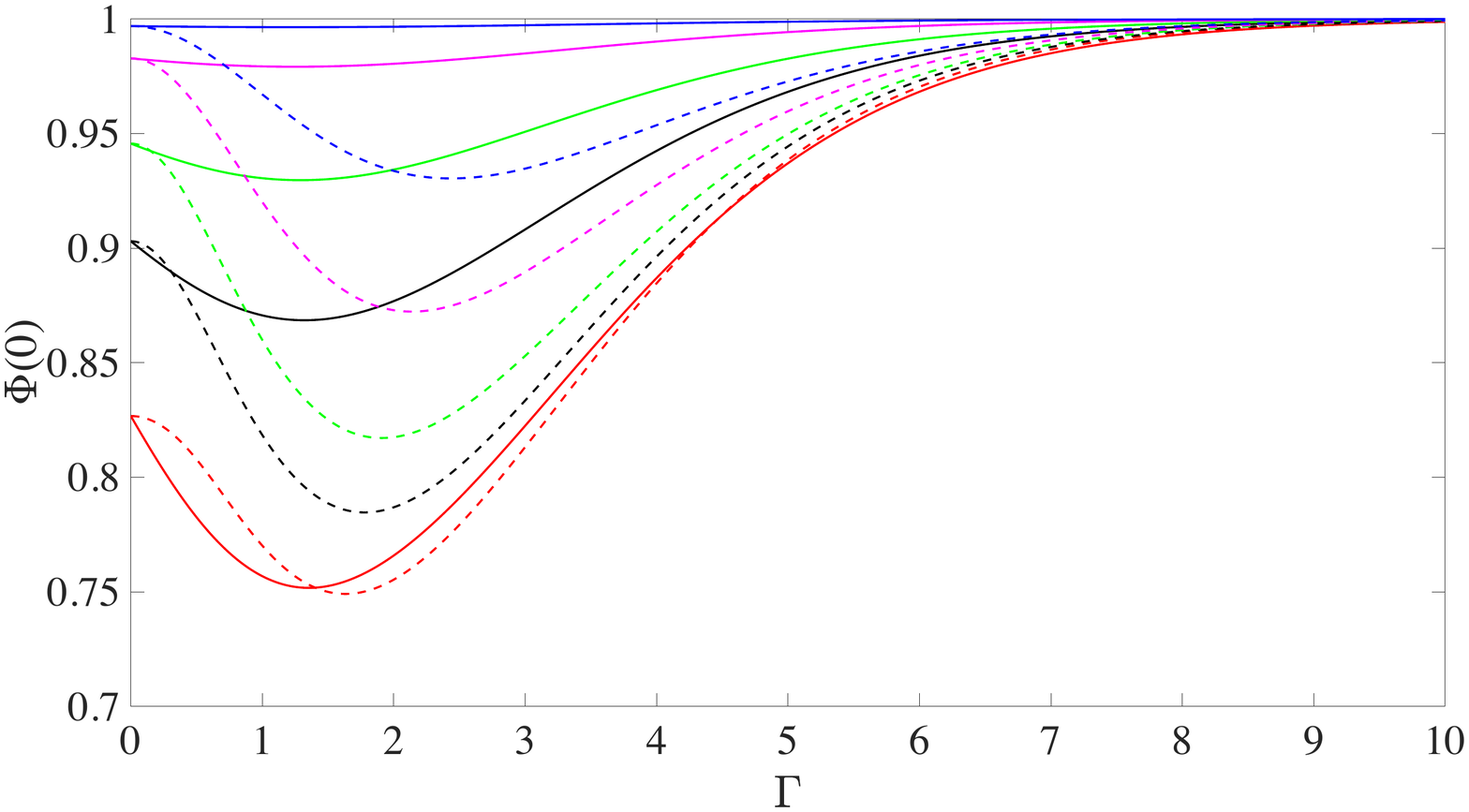}
	\caption{(Color online) Quantum yield as defined in Eq. (\ref{eqn:yield}) is plotted as a function of the strength of the confinement $\Gamma$. The vertical axis is made dimensionless by using $\Phi(0)=\tilde{\phi}(0)[\tau+\tau_e]/\tau \tau_e$ Every distance is expressed relative to the distance of the guest molecule to the focal point, i.e. $|x_e-x_c|$ and the strengths of the absorption at the trap and guest molecule are expressed through the dimensionless quantities $\mathcal{C}=C|x_e-x_c|/D$ and $\mathcal{C}_e=C_e|x_e-x_c|/D$, respectively. Starting from the bottom the dashed and solid curves, representing the reset and drift cases, respectively, are plotted with increasing values of $R$, namely $R=0.75$, 1, 1.25, 1.75 and 2.5. For all curves the trap and guest molecule absorption strength has been set equal with $\mathcal{C}=\mathcal{C}_e=4$.}
	\label{fig:yield} 
\end{figure}
To take into account the additional presence of the absorption site as well as the probability decay with rate $\tau_e$ the integral Eq. (\ref{eqn:FP_reset_traps}) is modified to
\bea
&&\fl \mathcal{P}_e(x,t|x_0,0)=e^{-\left(\frac{1}{\tau}+\frac{1}{\tau_e}\right)t}P_{\mbox{reset}}(x,t|x_0,0)\nonumber \\
&&-C\int_{0}^{t}ds\,
e^{-\left(\frac{1}{\tau}+\frac{1}{\tau_e}\right)(t-s)}P_{\mbox{reset}}(x,t-s|x',0)\mathcal{P}_e(x',s|x_0,0) \nonumber \\
&-&C_e\int_{0}^{t}ds\,
e^{-\left(\frac{1}{\tau}+\frac{1}{\tau_e}\right)(t-s)}P_{\mbox{reset}}(x,t-s|x',0)\mathcal{P}_e(x',s|x_0,0).
\label{eqn:FP_reset_traps_2}
\eea
Notice that both $\tau_e$ and $C_e$ may in principle differ from their respective counterparts radiative decay $\tau$ of the exciton moving through the host.

Once again the defect technique allows us to solve for $\mathcal{P}_e(x,t)$ in Eq. (\ref{eqn:FP-HO_exciton}) or (\ref{eqn:FP_reset_traps_2}). The problem is more complicated given that we are dealing with two defects rather than one, but the procedure is the same as for the one defect problem. As the solution of the defect-free problem with the addition of the loss term at rate $\mathcal{T}^{-1}=1/\tau+1/\tau_e$ is known, namely $\widetilde{P}\left(x,\epsilon+\mathcal{T}^{-1}|x_0,0\right)$, one can write the formal solution of Eq. (\ref{eqn:FP-HO_exciton}) as
\bea
\fl \widetilde{\mathcal{P}}_e\left(x,\epsilon|x_0,0\right)&=&\widetilde{P}\left(x,\epsilon+\mathcal{T}^{-1}|x_0,0\right)-C\widetilde{P}\left(x,\epsilon+\mathcal{T}^{-1}|x',0\right)\widetilde{\mathcal{P}}_e\left(x',\epsilon|x_0,0\right)
\nonumber \\
&-&C_e\widetilde{P}\left(x,\epsilon+\mathcal{T}^{-1}|x_e,0\right)\widetilde{\mathcal{P}}_e\left(x_e,\epsilon|x_0,0\right).
\label{eqn:formal_sol_2_def}
\eea
For Eq. (\ref{eqn:FP_reset_traps_2}) the solution is of the same form. It suffices to replace $\widetilde{P}(y,s|z,0)$ with $\widetilde{P}_{\mbox{reset}}(y,s|z,0)$.

To determine the general solution one needs to evaluate Eq. (\ref{eqn:formal_sol_2_def}) at the point $x'$ and $x_e$. One then finds $\widetilde{\mathcal{P}}_e\left(x_e,\epsilon|x_0,0\right)$ and $\widetilde{\mathcal{P}}_e\left(x',\epsilon|x_0,0\right)$ by solving two coupled algebraic equations. The result of the calculation gives the probability distribution and is shown in Appendix B. Using the explicit expression in Eq. (\ref{eqn:exact_sol_2_def}) it is now possible to calculate the `quantum yield'.  For the drift model it is given by
\bea
&&\fl \widetilde{\phi}(0)=\mathcal{T}\Bigg(1-\Bigg\{\left[1+C_e\widetilde{P}\left(x_e,\mathcal{T}^{-1}|x_e,0\right)\right]C\widetilde{P}\left(x',\mathcal{T}^{-1}|x_0,0\right)-C_e\widetilde{P}\left(x',\mathcal{T}^{-1}|x_e,0\right)\Bigg.\Bigg.\nonumber \\
&&\fl \times C\widetilde{P}\left(x_e,\mathcal{T}^{-1}|x_0,0\right)+\left[1+C\widetilde{P}\left(x',\mathcal{T}^{-1}|x',0\right)\right]C_e\widetilde{P}\left(x_e,\mathcal{T}^{-1}|x_0,0\right)-C\widetilde{P}\left(x_e,\mathcal{T}^{-1}|x',0\right) \nonumber \\
&&\fl \Bigg.\times C_e\widetilde{P}\left(x',\mathcal{T}^{-1}|x_0,0\right) \Bigg\}\Bigg\{\left[1+C\widetilde{P}\left(x',\mathcal{T}^{-1}|x',0\right)\right]\Bigg.\nonumber \\
&&\fl \left.\times
\Bigg.\left[1+C_e\widetilde{P}\left(x_e,\mathcal{T}^{-1}|x_e,0\right)\right]-C\widetilde{P}\left(x_e,\mathcal{T}^{-1}|x',0\right)C_e\widetilde{P}\left(x',\mathcal{T}^{-1}|x_e,0\right)\Bigg\}^{-1}\right),
\label{eqn:yield}
\eea
and with $\widetilde{P}_{\mbox{reset}}(y,s|z,0)$ replacing $\widetilde{P}(y,s|z,0)$ for the reset model.

In Fig. \ref{fig:yield}, where we plot $\widetilde{\phi}(0)$ as a function of the strength of the interaction $\Gamma=v|x_e-x_c|/D$ for the drift model and $\Gamma=\sqrt{r/D}|x_e-x_c|$ for the reset model, the non-monotonicity appears clearly and is more visible the larger is the dimensionless rate $R=|x_e-x_c|\sqrt{\mathcal{T}^{-1}/D}$. For the reset model the minimum of $\phi(0)$ shifts towards larger value as $\Gamma$ increases while it does not move appreciably for the drift model. This can be understood by noticing that $R$ is proportional with $\sqrt{\mathcal{T}}$. As $\Gamma$ is also proportional to a rate, i.e. $\sqrt{r}$, two decay processes may compete. As the minimum of the yield corresponds to when a particle has had the largest number of chances of being absorbed by the trap, an increase in $\mathcal{T}$ would limit those chances causing $\widetilde{\phi}(0)$ to increase across all values of $\Gamma$. But that decay competes with the decay $r$, while it does not in the drift model since the bias is deterministic and thus has no effect on the location of the minimum yield. 

For most values of $\Gamma$ the yield of the reset model is lower than the one for the drift model due to the weaker confinement of the resetting walker compared to the biased walker. This larger exploratory propensity, that is the tendency to stray further away from the focal point of the reset model in comparison to the drift model, as also indicated by the MSD (Fig. \ref{fig:msd}), is also responsible, as we will see in Sec. \ref{sec:msd_mfpt}, for the quantitative differences in the values of mean first-passage times.

\section{First-passage processes}
\label{sec:msd_mfpt}
In presence of a single trap at $x=x'$ we can determine the survival probability $M(t)$ by integrating the probability distribution at the trap site, given in Eq. (\ref{eqn:prop_defects}), over all space and obtain
\be
\widetilde{M}(\epsilon)=\frac{1}{\epsilon+\tau^{-1}}\left[1-\frac{\widetilde{P}(x',\epsilon+\tau^{-1}|x_0,0)}{1/C+\widetilde{P}(x',\epsilon+\tau^{-1}|x',0)}\right],
\label{eqn:survial_Laplace}
\ee
where it is assumed that $x_0\not=x'$.
In the limit $\tau^{-1}\rightarrow 0$ and $C\rightarrow +\infty$, that is for the case of a perfectly absorbing trap at $x'$ Eq. (\ref{eqn:survial_Laplace}) reduces to the Laplace expression of the relation between the first-passage probability distribution from $x_0$ to $x'$ and the survival probability, namely $-\epsilon \widetilde{M}_(\epsilon)+M(0)=\widetilde{\mathcal{F}}(x',\epsilon|x_0)$ with $M(0)=1$, or its time equivalent relation $-dM(t)/dt=\mathcal{F}(x',t|x_0)$.

The first-passage distribution $\mathcal{F}(x,t|x_0)$ from $x_0$ to $x$ for the two models can be expressed as
\be
\widetilde{\mathcal{F}}(x,\epsilon|x_0)=\frac{h_{x_c}(x,x_0)}{h_{x_c}(x,x)}\frac{\mathcal{C}(\epsilon)e^{-\alpha|x-x_0|\mathcal{S}(\epsilon)}+\mathcal{D}_{x_c}(x_0,\epsilon)e^{-\alpha|x-x_c|\mathcal{S}(\epsilon)}}{\mathcal{C}(\epsilon)+\mathcal{D}_{x_c}(x,\epsilon)e^{-\alpha|x-x_c|\mathcal{S}(\epsilon)}}
\label{eqn:first_passage_dist}
\ee
from which the mean first-passage time (MFPT) $\langle t_F(x|x_0)\rangle$ to reach $x$ from $x_0$ is obtained by performing $-\partial \mathcal{F}(x,\epsilon|x_0)/\partial \epsilon\Big|_{\epsilon=0}.$ After some algebra one can show that the general expression is
\bea
\fl \langle t_F(x|x_0)\rangle=&&\frac{h_{x_c}(x,x_0)}{h_{x_c}(x,x)}\frac{e^{\alpha|x-x_c|S(0)}}{\mathcal{D}^2_{x_c}(x,0)}\left\{\mathcal{C}'(0)\left[\mathcal{D}_{x_c}(x_0,0)-\mathcal{D}_{x_c}(x,0)e^{-\alpha|x-x_0|S(0)}\right]\right. \\ \nonumber
&&\left.+e^{-\alpha|x-x_c|\mathcal{S}(0)}\left[\mathcal{D}'_{x_c}(x,0)\mathcal{D}_{x_c}(x_0,0)-\mathcal{D}'_{x_c}(x_0,0)\mathcal{D}_{x_c}(x,0)\right]\right\}.
\label{eqn:mfpt_gen}
\eea
Using Eq. (\ref{eqn:cond_HO_lap}) the general expression above can be written explicitly for the constant drift model as
\be
\fl \langle t_F(x|x_0)\rangle=\frac{2D}{v^2}\left\{e^{\frac{v|x-x_c|}{D}}\left[1-e^{\frac{v(|x_0-x_c|-|x-x_0|-|x-x_c|)}{2D}}\right]+\frac{v(|x_0-x_c|-|x-x_c|)}{2D}\right\}.
\label{eqn:mfptV}
\ee
The form of Eq. (\ref{eqn:mfptV}) indicates that the time to reach a
target is composed of two contributions. For clarity of explanation, let
us suppose that $x_c<x<x_0$. In such a scenario, one finds that $\langle
t_F(x|x_0)\rangle=\frac{|x_0-x_c|-|x-x_c|}{v}$, that is, the mean first-passage time is simply the time it takes the mean position of the walker, initially at $x_0$, to reach $x_c$ moving at speed $v$. To consider the other case with $x>x_0$, let us first look at the case $x_0=x_c$, for which
\be
\langle t_F(x|x_c)\rangle=\frac{2D}{v^2}\left[e^{\frac{v|x-x_c|}{D}}-1-\frac{1}{2}\frac{v|x-x_c|}{D}\right].
\label{eqn:mfptV_origin}
\ee
With (\ref{eqn:mfptV_origin}), we can now see that the case in which
$x>x_0$ with $x,x_0>x_c$, the mean first-passage time is
\be
\langle t_F(x|x_0)\rangle=\langle t_F(x|
x_c)\rangle-\langle t_F(x_0|x_c)\rangle,
\label{eqn:mfptV-1}
\ee
that is, the average time one would need to get to $x$ from $x_c$
minus the contribution to get to $x_0$ from $x_c$. If $x_0<x_c$ but
$x>x_c$, then $\langle t_F(x|x_0)\rangle=\langle t_F(x|x_c)\rangle+\langle
t_F(x_c|x_0)\rangle$, that is, the contribution is the average time to get
to $x_c$ added to the average time to reach $x$ from $x_c$.

A natural comparison of the MFPT from $x_c$ to a
target at $x$ in presence of a potential may be made with the diffusive time between $x$ and $x_c$, that is $(x-x_c)^2/(2D)$. Denoting the ratio of the two by $g\left(\frac{v|x-x_c|}{D}\right)$, we have for the constant drift model
\be
g(z)=\frac{4}{z^2}\left(e^z-1-\frac{z}{2}\right).
\label{eqn:mfptV_norm-1}
\ee
For a fixed $|x-x_c|$ and $D$, as Eq. (\ref{eqn:mfptV_norm-1}) diverges both for $z\rightarrow 0$ and $z\rightarrow +\infty$, $g(z)$ possess a minimum. The divergence for small $z$ is expected as for $v\rightarrow 0$  the problem reduces to the unconstrained Brownian walker whose MFPT between any two points is infinite. The opposite limit, for very large $z$, that is when $v\rightarrow +\infty$, diverges because a point away from $x_c$ will never be reached. There is therefore an intermediate value of $v$ for which $g(z)$ is minimal. That value can be computed numerically from Eq. (\ref{eqn:mfptV_norm-1}) and equals $z^\star\simeq 1.245$. 

For the resetting case, using the expressions in (\ref{eqn:cond_reset_lap}) in Eq. (\ref{eqn:mfpt_gen}), it follows that the MFPT
\be
\langle t_F(x|x_0)\rangle=\frac{e^{|x-x_c|\sqrt{r/D}}}{r}\Big(1-e^{-|x-x_0|\sqrt{r/D}}\Big).
\label{eqn:MFPTdist_reset}
\ee
The dimensionless expression
$g\left(|x-x_c|\sqrt{r/D}\right)=2D\langle t_F(x|x_c)\rangle /(x-x_c)^2$ that
represents a walker  reaching $x$ for the first time while starting from
$x_0=x_c$ can be written as 
\be
g(z)=2\frac{e^z-1}{z^2}.
\label{eqn:MFPT_reset_dimensionless}
\ee
Similarly to the drift model for a given fixed $|x-x_c|$ and a given diffusion constant, the above
expression implies a finite MFPT at a finite $r$, and a divergence at
the two limits $r \to 0$ (pure diffusion) and $r \to +\infty$. In the
latter case, too frequent resets for the motion to reach the target in between lead to an infinite MFPT. In fact, as a function of $r$, the MFPT has its minimum at an optimal resetting rate $r_{\rm opt}=(z^\star)^2D/|x-x_0|^2$, where $z^\star$ solves the transcendental equation $z^\star/2=1-e^{-z^\star}$, and has the numerical value $z^\star=1.59362...$ \cite{evansmajumdar2011b}. 

\begin{figure}[!h]
\includegraphics[width=150mm]{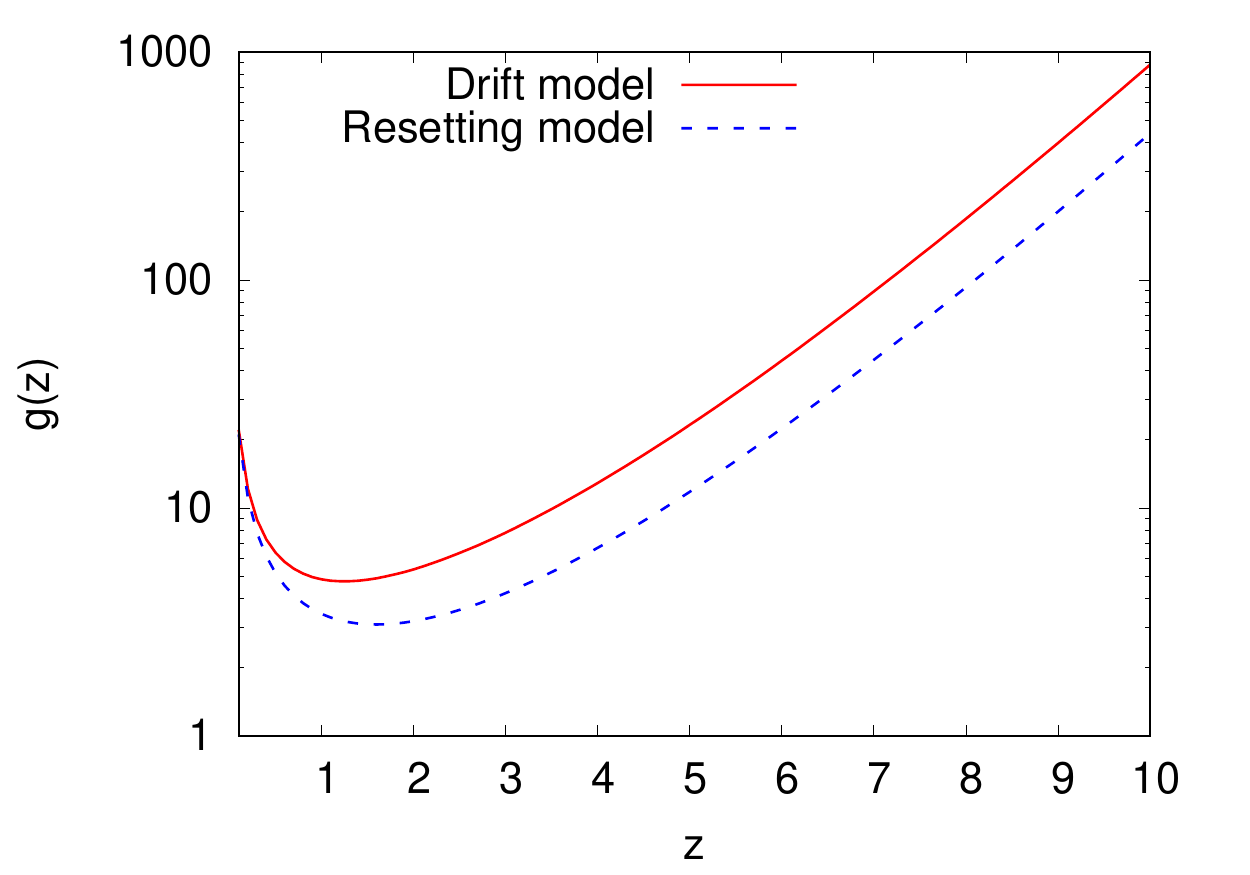}
\caption{(Color online) Comparison of the dimensionless MFPT for the constant drift model and the resetting model. Here, $z=\frac{v|x-x_c|}{D}$ for the drift model, while $z=\frac{|x-x_c|\sqrt{r}}{\sqrt{D}}$ for the resetting model. The function $g(z)$ is given by Eq.  (\ref{eqn:mfptV_norm-1}) for the drift model and by Eq. (\ref{eqn:MFPT_reset_dimensionless}) for the resetting model. }
\label{fig:mfpt-comparison}
\end{figure}

The comparison of Eqs.  (\ref{eqn:mfptV_norm-1})  and
(\ref{eqn:MFPT_reset_dimensionless}) for the dimensionless $g(z)$
displays the minima for both models. The plot in Fig.
\ref{fig:mfpt-comparison} also indicates that the MFPT from $x_c$ to any
point in space for the drift model is always larger than the
corresponding one for the reset model. As mentioned already in
presenting the quantum yield in Sec. \ref{sec:traps}, this is the result
of the comparatively weaker tendency of the reset model to return to
$x_c$. Note that the results summarized in Fig.~\ref{fig:mfpt-comparison}
have already been reported earlier in Ref.~\cite{evansmajumdarmallick2013} by the
approach of backward Fokker-Planck equation.

\section{Splitting probability and propagators in presence of two absorbing targets}
\label{sec:split}

A natural extension of the case $C\rightarrow +\infty$ for one trap is the problem of two absorbing defects. It corresponds to Eqs. (\ref{eqn:FP-HO_exciton}) and (\ref{eqn:FP_reset_traps_2}) with $C$, $C_e$, $\tau$ and $\tau_e$ infinitely large. As the method of images can only be applied when the space is homogeneous, the presence of a focal point forces us to employ a different methodology. We analyze in particular the situation when one defect is at $x=a$ and the other at
$x=b$, respectively, to the left and to the right of the focal point (for simplicity $x_c=0$), and with the initial location of the walker being
$a<x_0<b$. Following Ref. \cite{giornoetal2011}, we proceed by constructing the
general expression for the propagator in presence of these absorbing
boundaries. We start first with the case of a single absorbing boundary at
$x=a$. One takes the propagator without boundaries $P(x,t|x_0)$
and constructs the one in presence of the absorbing boundary by subtracting all the
trajectories that go through $x=a$ at some time $t'<t$. Denoting by $Q_a(x,t|x_0)$ the distribution for the
ensemble of such trajectories, we can determine it by relating it to $P(x,t|x_0,0)$ and $P(x,t|a,0)$
via the first-passage probability $\mathcal{F}(a,t|x_0)$ to be at $x$ at some time $t$
starting at $x_0$. In fact, we have that $\mathcal{F}(a,s|x_0)P(x,t-s|a)$
represents the probability distribution of having reached $x=a$ at time
$s$ (earlier than $t$) and subsequently having reached $x$ at time $t$
starting from $x=a$. Integrating up to time $t$ gives the relation
$H_a(x,t|x_0)=\int_{0}^{t}{\rm d}s~\mathcal{F}(a,s|x_0)P(x,t-s|a)$. If
we now subtract $H_a(x,t|x_0)$ from $P(x,t|x_0)$, we have the
probability of all trajectories that are at $x$ at time $t$ that have
not reached $a$ at an earlier time. In Laplace space, we simply write \be
\widetilde{\mathcal{A}}_a(x,\epsilon|x_0)=\widetilde{P}(x,\epsilon|x_0)-\widetilde{\mathcal{F}}(a,\epsilon|x_0)\widetilde{P}(x,\epsilon|a),
\label{eqn:prop_1absorb}
\ee
which is the Laplace propagator for the walker being at $x$ at time $t$ starting at $x_0$ in presence of an absorbing boundary at $x=a$.

We now consider the other absorbing boundary at $x=b$ and we determine the probability of having reached $b$ without having gone through $x=a$ and vice versa. Let us call $\mathcal{G}_a(b,t|x_0)$ the probability density of having reached $b$ at time $t$ without having been at $x=a$ at an earlier time. We can now use the propagator $\mathcal{A}_a(x,t|x_0)$ to do it since we use the relation
\be
\mathcal{A}_a(b,t|x_0)=\int_{0}^{t}{\rm d}s\,\mathcal{G}_a(b,s|x_0)\mathcal{A}_a(b,t-s|b).
\label{eqn:prop_2absorb}
\ee
Eq. (\ref{eqn:prop_2absorb}) is analogous to what used earlier to define a relation between first-passage at $x=b$ and
 probability distribution to be at $x$ at time $t$. That general relation is here employed with the propagator in presence of an absorbing boundary at $a$ in place of the usual
 propagator $P$ in the absence of any boundary. In Laplace it is simply
 $\widetilde{\mathcal{A}}_a(b,\epsilon|x_0)=\widetilde{\mathcal{G}}_a(b,\epsilon|x_0)\widetilde{\mathcal{A}}_a(b,\epsilon|b)$. We can thus write
 \be
 \widetilde{\mathcal{G}}_a(b,\epsilon|x_0)=\frac{\widetilde{\mathcal{A}}_a(b,\epsilon|x_0)}{\widetilde{\mathcal{A}}_a(b,\epsilon|b)}.
\label{eqn:tdep_split_prob}
\ee
We can similarly construct the case in which we consider first the absorbing boundary at $x=b$ for which we write the propagator as  
\be
\widetilde{\mathcal{A}}_b(x,\epsilon|x_0)=\widetilde{P}(x,\epsilon|x_0)-\widetilde{\mathcal{F}}(b,\epsilon|x_0)\widetilde{P}(x,\epsilon|b),
\label{eqn:a_b}
\ee
and similarly
\be
\widetilde{\mathcal{G}}_b(a,\epsilon|x_0)=\frac{\widetilde{\mathcal{A}}_b(a,\epsilon|x_0)}{\widetilde{\mathcal{A}}_b(a,\epsilon|a)}.
\label{eqn:split_ba}
\ee
 In extended form we have that
\be
\widetilde{\mathcal{G}}_a(b,\epsilon|x_0)=\frac{\widetilde{P}(b,\epsilon|x_0)-\widetilde{\mathcal{F}}(a,\epsilon|x_0)\widetilde{P}(b,\epsilon|a)}{\widetilde{P}(b,\epsilon|b)-\widetilde{\mathcal{F}}(a,\epsilon|b)\widetilde{P}(b,\epsilon|a)},
\label{eqn:gab_eps}
\ee
and
\be
\widetilde{\mathcal{G}}_b(a,\epsilon|x_0)=\frac{\widetilde{P}(a,\epsilon|x_0)-\widetilde{\mathcal{F}}(b,\epsilon|x_0)\widetilde{P}(a,\epsilon|b)}{\widetilde{P}(a,\epsilon|a)-\widetilde{\mathcal{F}}(b,\epsilon|a)\widetilde{P}(a,\epsilon|b)},
\label{eqn:gba_eps}
\ee
\begin{figure}[!h]
\includegraphics[width=150mm]{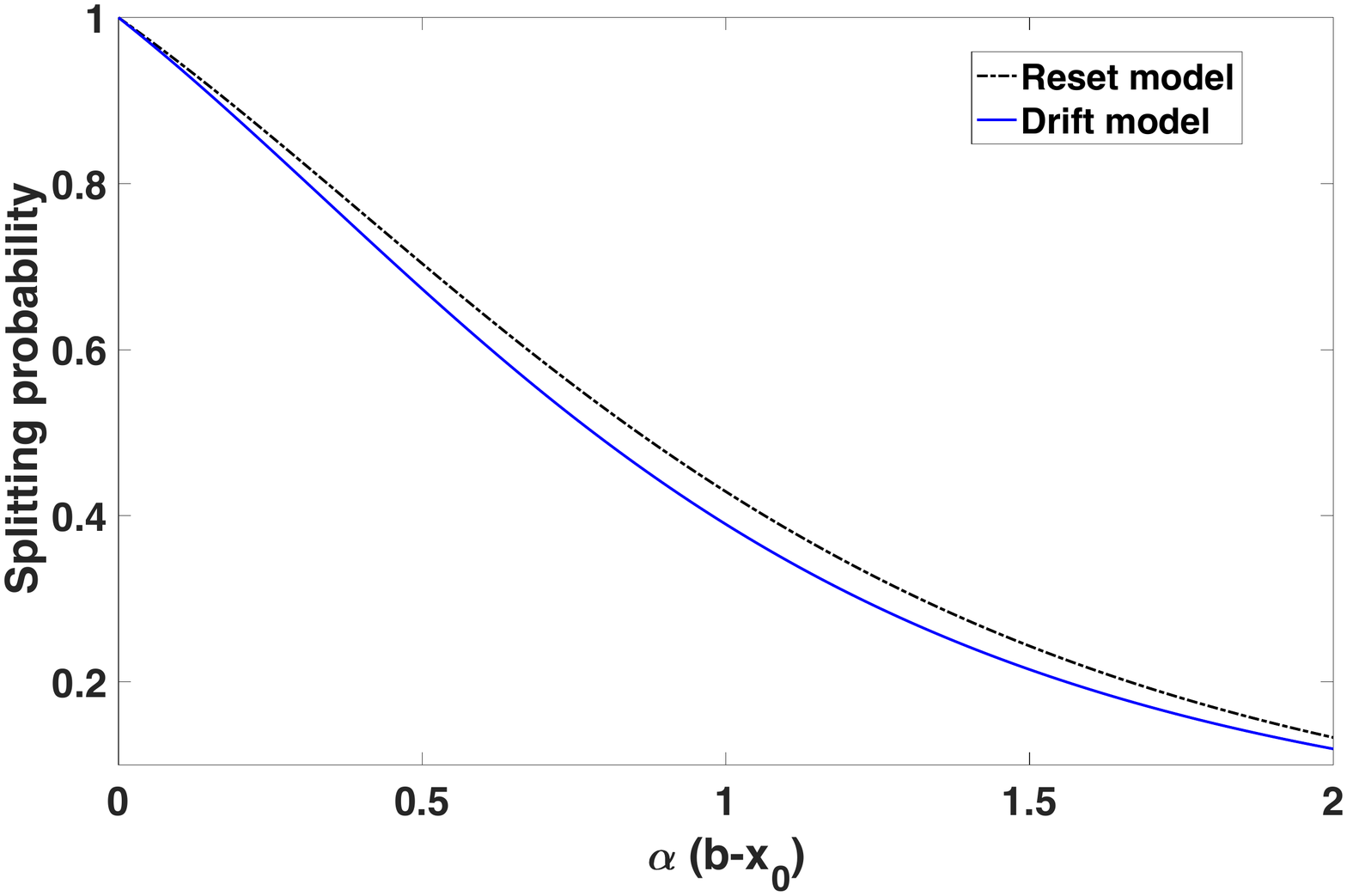}
\caption{(Color online) Comparison of the splitting probability $\widetilde{\mathcal{G}}_a(b,0|x_0)$ for the two models, namely the probability of reaching $x=b$ without ever reaching $x=a$. The focal point is set at $\alpha x_c=0$, the initial condition at $\alpha x_0=2$ and $\alpha |b-a|=4$. The data is generated by a rigid translation of $a$ and $b$ to the right in Eqs. (\ref{eqn:split_prob_vpot}) and (\ref{eqn:split_prob_exp_reset}) for the drift and reset model, respectively. The higher splitting probability for the reset model is another indication of the stronger exploratory nature of its walk compared to the drift model.}
\label{fig:split}
\end{figure}
with the latter expression also derivable form the former by a simple exchange of $a$ and $b$. While we have reported the explicit form of Eq. (\ref{eqn:gab_eps}) in Appendix C in Eq. (\ref{eqn:full_split_prob}), we write here explicitly the 
quantities $\widetilde{\mathcal{G}}_a(b,0|x_0)$
and $\widetilde{\mathcal{G}}_b(a,0|x_0)$ (notice that
$\widetilde{\mathcal{G}}_a(b,0|x_0)+\widetilde{\mathcal{G}}_b(a,0|x_0)=1$), which are the so-called splitting probabilities, respectively, of reaching $b$ without having reached $a$, and of reaching $a$ without having reached $b$. For the drift model we have
\be
\fl \widetilde{\mathcal{G}}_a(b,0|x_0)=
\frac{e^{-\frac{v|b|}{D}}-e^{-\frac{v(|b|+|a|+|b-a|)}{2D}}+e^{-\frac{v(|b|+2|a|+|b-x_0|-|x_0|)}{2D}}-e^{-\frac{v(|a|+2|b|+|a-x_0|-|x_0|)}{2D}}}{e^{-\frac{v|a|}{D}}+e^{-\frac{v|b|}{D}}-2e^{-\frac{v(|a|+|b|+|b-a|}{2D}}},
\label{eqn:split_prob_vpot}
\ee
while for the reset model we obtain the expression
\be
\fl \widetilde{\mathcal{G}}_a(b,0|x_0)=\frac{e^{|b-a|\sqrt{r/D}}\left(e^{(|b|-|b-x_0|)\sqrt{r/D}}-e^{(|a|-|a-x_0|)\sqrt{r/D}}\right)+e^{(|a|+|b-a|)\sqrt{r/D}}-e^{|b|\sqrt{r/D}}}{\left(e^{|b-a|\sqrt{r/D}}-1\right)\left(e^{|a|\sqrt{r/D}}+e^{|b|\sqrt{r/D}}\right)}.
\label{eqn:split_prob_exp_reset}
\ee
\begin{figure}[!h]
\includegraphics[width=160mm]{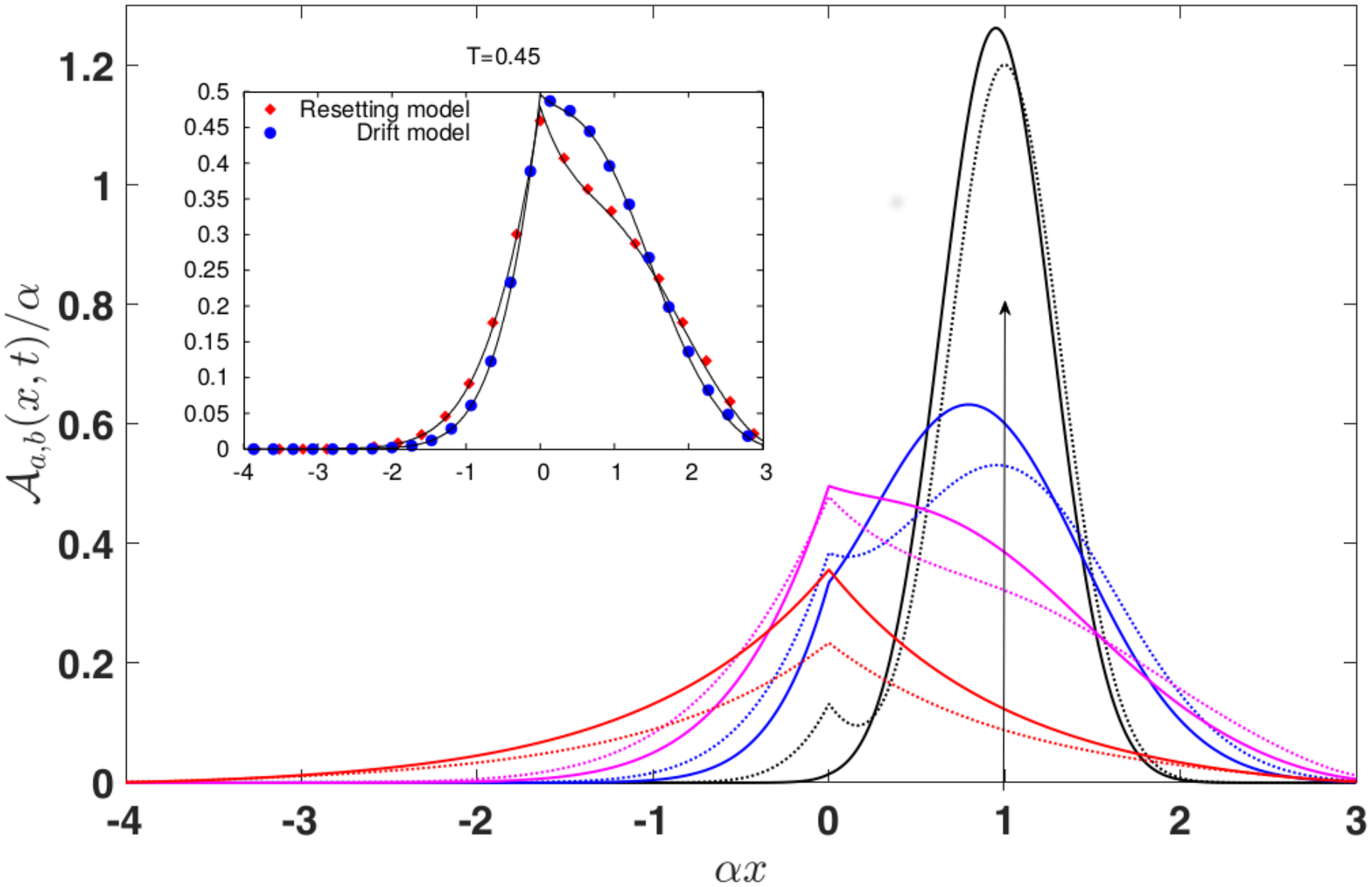}
\caption{(Color online) Comparison of the spatio-temporal dynamics of
        $\mathcal{A}_{a,b}(x,t)$ with $x_c=0$ for the the constant drift
        model (solid lines) and the reset model (dashed lines) in
        presence of two absorbing boundaries at $x=a$ and $x=b$. The
        Dirac delta initial condition, indicated by the arrow, is
        located at $\alpha x_0=1$ and the left and right boundaries are,
        respectively, at $\alpha a=-4$ and $\alpha b=3$. Starting from
        the highest curves on the right, the different lines for the two
        models are plotted at times $T=0.05$, $0.2$, $0.45$, and $9.95$. The
        inset shows the match between theory and simulations for
        $T=0.45$.}
\label{fig:abs_prob}
\end{figure}

Analogously to the differences observed for the MSD, the quantum yield and the MFPT analysis, we expect the splitting probability to be higher for the reset model compared to the drift model because a walker subject to a deterministic drift is more likely to roam closer to $x_c$ and thus reduce the chance of encountering the forbidden target. This difference is shown in Fig. \ref{fig:split} where we plot the splitting probability $\widetilde{\mathcal{G}}_a(b,0|x_0)$.

To determine the general
propagator $\widetilde{\mathcal{A}}_{a,b}(x,\epsilon|x_0)$ in presence of an absorbing boundary at $x=a$ and $x=b$ it is necessary to subtract from the free propagation the contribution of those trajectories that reach first $x=a$ before reaching $x=b$ and those that reach $x=b$ before reaching $x=a$. The resulting propagator is
\be
\widetilde{\mathcal{A}}_{a,b}(x,\epsilon|x_0)=\widetilde{P}(x,\epsilon|x_0)-\widetilde{\mathcal{G}}_a(b,\epsilon|x_0)\widetilde{P}(x,\epsilon|b)-\widetilde{\mathcal{G}}_b(a,\epsilon|x_0)\widetilde{P}(x,\epsilon|a),
\label{eqn:propag_2bound}
\ee
which by construction is identically zero at $x=a$ and $x=b$.
The dynamics of $\mathcal{A}_{a,b}(x,t|x_0)$ for the two models is displayed in Fig. \ref{fig:abs_prob}.

\section{Conclusions}
\label{sec:concl}

We have presented a systematic analysis of two models of tethered motion. The analytic knowledge of the propagators for the two processes has been instrumental in our findings. We have shown how the fundamental difference of the two models, that is, the biased movement through the intervening space for the drift model versus the random and sudden jumps to the focal point for the resetting model, affects the dynamics of the walker. We have demonstrated the effects by not only comparing the spatio-temporal dynamics of the propagators, but also deriving quantities such as the mean first-passage time, splitting probabilities, and the probability distribution in presence of absorbing boundaries.

Compared to the resetting model, the deterministic nature of the bias in the drift model limits the spatial extent over which the walker moves away from the focal point. In other words the two models exhibit a different exploratory dynamics around the focal point. As the possibility for larger exploration favors the walker to cover the available space, search for a randomly located item is more efficient with the reset model, as shown by the analysis of the mean first-passage time. For the same reason, a better performance of the reset model for random search of multiple targets is also expected, as shown by our comparison of the splitting probabilities of perfectly absorbing targets. Calculations for the quantum yield with a guest molecule and a trap, both away from the focal point, shows that the drift model for different choices of absorption reaches higher values for most cases of confinement strength because the absorbing targets are reached less often when compared to the reset model.

The formalism developed here for the drift model has applications in studying animal movement where the model is well known since the '70s, \cite{holgate1971}, under the name Holgate-Okubo model \cite{okubolevinbook2001}. It was introduced in the literature to model many examples of animal roaming within their own home range and for which a focal point represents a burrow or a den site, while the tendency to return is represented by biasing the random movement with a drift. Despite its long history, it is only recently that the derivation of an analytic expression for the propagator has been achieved, first in 2010 \cite{touchetteetal2010} and, through a different technique, in 2016 \cite{chaseetal2016} by some of the present authors.

Despite its simplicity, the Holgate-Okubo model has been extensively
employed to model confined animal movement, and many variants have appeared over the years. A simple way to introduce variation to the model is to change the 
confining potential---proportional to the modulus of the distance from the focal point in the drift model---to a shallower or steeper spatial profile \cite{giuggiolietal2006}. One example is the use of a quadratic
potential, whereby the Fokker-Planck equation of the drift model reduces to that of the Ornstein-Uhlenbeck process
\cite{uhlenbeckornstein1930} representing the noisy, overdamped dynamics of a
walker being pulled towards the focal point by a harmonic spring.

The history of the stochastic resetting reset model is instead much more recent. It has emerged in recent years as an interesting and an analytically
tractable paradigm to study stationary
processes out of equilibrium. A variety of situations have been considered, e.g., a diffusing particle resetting to its initial position in either a free
\cite{evansmajumdar2011a,evansmajumdar2014,majumdarsabhapanditschehr20151,reuveni2016,eulemetzger2016,nagargupta2015},
or a bounded domain \cite{christou2015}, in presence of an external
potential \cite{pal2015}, and for different choices
of the resetting position
\cite{evansmajumdar2011b,boyersolissalas2014,majumdarsabhapanditschehr20152}.
Other generalizations include resetting of continuous-time random walks \cite{monterovillarroel2013,mendezcampos2016}, L\'{e}vy \cite{kusmierzmajumdarsabhapanditschehr2014} and
exponential constant-speed flights \cite{camposmendez2015}, and
time-dependent resetting of a Brownian particle \cite{palkunduevans2016}.
Stochastic resetting has also been invoked in the context of
reaction-diffusion models \cite{duranghenkelpark2014}, fluctuating
interfaces \cite{guptamajumdarschehr2014,guptanagar2016}, in modeling
backtrack recovery by RNA polymerases \cite{roldanetal2016}, and in
discussing stochastic thermodynamics far from equilibrium
\cite{fuchsgoldtseifert2016}. Various static and time-dependent aspects of
resetting dynamics have been studied over the years, e.g., in discussing
first-passage properties \cite{Pal:2017}, optimal search strategies
\cite{Evans:2013,Kusmierz:2017,Belan:2018} and first-arrival times
\cite{Kusmierz:2015}, and resetting in a bounded
domain \cite{Christou:2015,Chatterjee:2018}. A recent study of resetting, invoking the path-integral formalism of stochastic processes, has shown how resetting may have practical applications in confining diffusing particles in space. For example, it was demonstrated in the case of energy-dependent resetting in presence of a harmonic potential that a specified degree of spatial confinement of Brownian
particles is possible on a much faster time scale than what may be achieved by performing quenches of parameters of the harmonic potential \cite{roldan2017}. Other possible applications of our reset formalism could be in discussing the contracted movement of a random walker from a higher dimensional space with effective long range jumps. A typical example would be the diffusion onto a DNA sequence \cite{benichouetal2009}. On one hand, two bases might be very far from each other, if some substance has to be transported along the sequence. On the other hand, the two base may be considered just a single step away if the coiling of the DNA in three dimensions makes the two base touch each other.
We wrap off by mentioning future directions to pursue that could include extending our analysis to consider the presence of reflecting boundary conditions, dynamics in higher dimensions, as well as the case of many interacting random walkers. 

 \renewcommand{\theequation}{A\arabic{equation}}    
  \setcounter{equation}{0} 
\section*{Appendix A}
\label{sec:appA}
For completeness we report the MSD expression for the drift model for the general case with $x_0$ and $x_c$ not identically zero. The calculation proceeds in two steps given that the propagator in Eq. (\ref{eqn:time-HO}) can be expressed as the sum of two terms $P_1(x,t)$ and $P_2(x,t)$, the latter being the term containing the error function. For the first term one needs to integrate
\bea
&& \fl I_1(t)=\int_{-\infty}^{+\infty}dx(x-x_0)^2P_1(x,t) \nonumber \\
&&\fl =F(x_0,x_c,t) \int_{0}^{+\infty}dy\,e^{-\frac{y^2}{4Dt}-\frac{vy}{2D}}\left\{\left[2\gamma^2+2y^2\right]\cosh\left(\frac{y\gamma}{2Dt}\right) -4\gamma y \sinh\left(\frac{y\gamma}{2Dt}\right)\right\},
\label{eqn:msd-HO_1}
\eea
with $F(x_0,x_c,t)=\frac{1}{4\pi Dt}e^{\frac{v|x_c-x_0|}{2D}-\frac{v^2t}{4D}-\frac{(x_c-x_0)^2}{4Dt}}$ and $\gamma=x_c-x_0$, whereas for the second term the integration reduces to the calculation of the integral
\bea
I_2(t)&=&\int_{-\infty}^{+\infty}dx\,(x-x_0)^2P_2(x,t)  \nonumber \\
&=&\frac{v}{2}\sqrt{\frac{t}{D}}\int_{0}^{+\infty}dy\,e^{-2yv\sqrt{\frac{t}{D}}}\,(4Dty^2+\gamma^2)\,{\rm Erfc}\left(y+\frac{|\gamma|-vt}{\sqrt{4Dt}}\right).
\label{eqn:msd-HO_2}
\eea
Performing the integrations give
\bea
&&\fl I_1(t)=-\frac{2}{\sqrt{\pi}}\sqrt{Dt}\,vt\,e^{-\frac{1}{4}\left(\sqrt{\frac{t}{D}}v-\frac{|\gamma|}{\sqrt{Dt}}\right)^2} +\left(Dt+\frac{v^2t^2}{2}\right)\nonumber \\
&&\fl \times\left\{e^{\frac{v(|\gamma|-\gamma)}{2D}}{\rm Erfc}\left[\frac{1}{2}\left(\sqrt{\frac{t}{D}}v-\frac{\gamma}{\sqrt{Dt}}\right)\right]+e^{\frac{v(|\gamma|+\gamma)}{2D}}{\rm Erfc}\left[\frac{1}{2}\left(\sqrt{\frac{t}{D}}v+\frac{\gamma}{\sqrt{Dt}}\right)\right]\right\},
\label{eqn:I1}
\eea
and
\bea
&&\fl I_2(t)=2\frac{D^2}{v^2}+\gamma^2+\left[\sqrt{\frac{Dt}{\pi}}\left(|\gamma|+vt-2\frac{D}{v}\right)\right]e^{-\frac{1}{4}\left(\sqrt{\frac{t}{D}}v-\frac{|\gamma|}{\sqrt{Dt}}\right)^2}-{\rm Erfc}\left[\frac{1}{2}\left(\sqrt{\frac{t}{D}}v-\frac{|\gamma|}{\sqrt{Dt}}\right)\right] \nonumber \\
&&\fl
\times\left(\frac{D^2}{v^2}+\frac{\gamma^2}{2}\right)-e^{\frac{v|\gamma|}{D}}\left\{vt|\gamma|+\frac{D^2}{v^2}\left(1-\frac{v|\gamma|}{D}\right)+\frac{v^2t^2}{2}+\frac{\gamma^2}{2}\right\}{\rm
Erfc}\left[\frac{1}{2}\left(\sqrt{\frac{t}{D}}v+\frac{|\gamma|}{\sqrt{Dt}}\right)\right].
\nonumber \\
\label{eqn:I2}
\eea
Combining Eq. (\ref{eqn:I1}) and (\ref{eqn:I2}) one obtains
\bea
&&\fl \langle (x-x_0)^2\rangle(t)=2\frac{D^2}{v^2}+\gamma^2+\left[\sqrt{\frac{Dt}{\pi}}\left(|\gamma|-vt-2\frac{D}{v}\right)\right]e^{-\frac{1}{4}\left(\sqrt{\frac{t}{D}}v-\frac{|\gamma|}{\sqrt{Dt}}\right)^2} \nonumber \\
&&\fl +\left(Dt+\frac{v^2t^2}{2}-\frac{D^2}{v^2}-\frac{\gamma^2}{2}\right){\rm Erfc}\left[\frac{1}{2}\left(\sqrt{\frac{t}{D}}v-\frac{|\gamma|}{\sqrt{Dt}}\right)\right]  \nonumber \\
&&\fl +e^{\frac{v|\gamma|}{D}}\left\{Dt-vt|\gamma|-\frac{D^2}{v^2}\left(1-\frac{v|\gamma|}{D}\right)-\frac{\gamma^2}{2}\right\}{\rm Erfc}\left[\frac{1}{2}\left(\sqrt{\frac{t}{D}}v+\frac{|\gamma|}{\sqrt{Dt}}\right)\right],
\label{eqn:I1plusI2}
\eea
which reduces to the one reported in the main text in Eq. (\ref{eqn:msd_centred}).

 \renewcommand{\theequation}{B\arabic{equation}}    
  \setcounter{equation}{0} 
\section*{Appendix B}
\label{sec:appB}
To determine the exact form of $\widetilde{\mathcal{P}}_e\left(x_e,\epsilon|x_0,0\right)$ and $\widetilde{\mathcal{P}}_e\left(x',\epsilon|x_0,0\right)$  in Sec. \ref{sec:traps} we assign $x=x'$ and $x=x_e$ to $\widetilde{\mathcal{P}}_e\left(x,\epsilon+\mathcal{T}^{-1}|x_0,0\right)$ in Eq. (\ref{eqn:formal_sol_2_def}) and we are left to solve the matricial equation
\bea
&&\fl \left(\begin{array}{cc} 1+C\widetilde{P}\left(x',\epsilon+\mathcal{T}^{-1}|x',0\right) & C_e \widetilde{P}\left(x',\epsilon+\mathcal{T}^{-1}|x_e,0\right) \\ C\widetilde{P}\left(x_e,\epsilon+\mathcal{T}^{-1}|x',0\right) & 1+C_e\widetilde{P}\left(x_e,\epsilon+\mathcal{T}^{-1}|x_e,0\right) \end{array}\right)\left(\begin{array}{c} \widetilde{\mathcal{P}}_e\left(x',\epsilon|x_0,0\right)\\ \widetilde{\mathcal{P}}_e\left(x_e,\epsilon|x_0,0\right) \end{array}\right) \nonumber \\
&&\fl=\left(\begin{array}{c} \widetilde{P}\left(x',\epsilon+\mathcal{T}^{-1}|x_0,0\right)\\ \widetilde{P}\left(x_e,\epsilon+\mathcal{T}^{-1}|x_0,0\right) \end{array}\right).
\label{eqn:matricial}
\eea
Inverting the matrix allows us to determine $\widetilde{\mathcal{P}}_e\left(x_e,\epsilon|x_0,0\right)$ and $\widetilde{\mathcal{P}}_e\left(x',\epsilon|x_0,0\right)$  and thus obtain the general expression for $\widetilde{\mathcal{P}}_e\left(x,\epsilon|x_0,0\right)$ in Eq. (\ref{eqn:formal_sol_2_def}), namely
\bea
&&\fl \widetilde{\mathcal{P}}_e\left(x,\epsilon|x_0,0\right)=\widetilde{P}\left(x,\epsilon+\mathcal{T}^{-1}|x_0,0\right)-C\widetilde{P}\left(x,\epsilon+\mathcal{T}^{-1}|x',0\right)\left\{\left[1+C_e\widetilde{P}\left(x_e,\epsilon+\mathcal{T}^{-1}|x_e,0\right)\right]\right.
\nonumber \\
&&\fl \left.\times\widetilde{P}\left(x',\epsilon+\mathcal{T}^{-1}|x_0,0\right)-C_e \widetilde{P}\left(x',\epsilon+\mathcal{T}^{-1}|x_e,0\right)\widetilde{P}\left(x_e,\epsilon+\mathcal{T}^{-1}|x_0,0\right)\right\}\mathcal{A}(\epsilon) \nonumber \\
&&\fl -C_e\widetilde{P}\left(x,\epsilon+\mathcal{T}^{-1}|x_e,0\right)\left\{\left[1+C\widetilde{P}\left(x',\epsilon+\mathcal{T}^{-1}|x',0\right)\right]C\widetilde{P}\left(x_e,\epsilon+\mathcal{T}^{-1}|x_0,0\right)\right. \nonumber \\
&&\fl -\left.C\widetilde{P}\left(x_e,\epsilon+\mathcal{T}^{-1}|x',0\right)\widetilde{P}\left(x',\epsilon+\mathcal{T}^{-1}|x_0,0\right)\right\}\mathcal{A}(\epsilon),
\label{eqn:exact_sol_2_def}
\eea
where
\bea
\mathcal{A}(\epsilon)&=&\left\{\left[1+C\widetilde{P}\left(x',\epsilon+\mathcal{T}^{-1}|x',0\right)\right]\left[1+C_e\widetilde{P}\left(x_e,\epsilon+\mathcal{T}^{-1}|x_e,0\right)\right]\right. .\nonumber \\
&-&\left.C\widetilde{P}\left(x_e,\epsilon+\mathcal{T}^{-1}|x',0\right)C_e\widetilde{P}\left(x',\epsilon+\mathcal{T}^{-1}|x_e,0\right)\right\}^{-1}.
\label{eqn:det_transf}
\eea
In the absence of traps or guest molecule and the associated probability decay, that is respectively with $C_e\rightarrow 0$ and $\tau_e\rightarrow +\infty$ or $C\rightarrow 0$ and $\tau\rightarrow +\infty$, the form of the propagator reduces to the one found in Eq. (\ref{eqn:prop_defects}).

 \renewcommand{\theequation}{C\arabic{equation}}    
  \setcounter{equation}{0} 
\section*{Appendix C}
\label{sec:appC}
Here we explicitly show the Laplace expression (\ref{eqn:tdep_split_prob}) for the splitting probability distribution $\mathcal{G}_a(b,\epsilon|x_0)$, that is the probability distribution of reaching $x=b$ without having reached $x=a$. With the notation explained in Sec. \ref{sec:models} it is given by 
\bea 
&&\fl \mathcal{G}_a(b,\epsilon|x_0)=h_{x_c}(b,x_0)\left\{\mathcal{C}(\epsilon)\left[e^{-\alpha|b-x_0|\mathcal{S}(\epsilon)}-e^{-\alpha(|a-x_0|+|a-b|)\mathcal{S}(\epsilon)}\right]+\mathcal{D}_{x_c}(a,\epsilon)\left[e^{-\alpha(|b-x_0|+|a|)\mathcal{S}(\epsilon)}\right.\right. \nonumber \\
&&\fl \left.\left.-e^{-\alpha(|a-x_0|+|b|)\mathcal{S}(\epsilon)}\right]+\mathcal{D}_{x_c}(x_0,\epsilon)\left[e^{-\alpha|b|\mathcal{S}(\epsilon)}-e^{-\alpha(|a-b|+|a|)\mathcal{S}(\epsilon)}\right]\right\}\left\{\mathcal{C}(\epsilon)\left[1-e^{-2\alpha|a-b|\mathcal{S}(\epsilon)}\right]\right.\nonumber \\
&&\fl +\mathcal{D}_{x_c}(a,\epsilon)\left[e^{-\alpha|a|\mathcal{S}(\epsilon)}-e^{-\alpha(|a-b|+|b||)\mathcal{S}(\epsilon)}\right]+\mathcal{D}_{x_c}(b,\epsilon)\left[e^{-\alpha|b|\mathcal{S}(\epsilon)}-e^{-\alpha(|a-b|+|a|)\mathcal{S}(\epsilon)}\right].
\label{eqn:full_split_prob}
\eea
It is this expression for $\mathcal{G}_a(b,\epsilon|x_0)$ and the corresponding $\mathcal{G}_b(a,\epsilon|x_0)$ that we have Laplace inverted and then convoluted, respectively, with the known time-dependent functions $\tilde{P}(x,\epsilon|b)$ and $\tilde{P}(x,\epsilon|a)$ to reconstruct the probability distribution and plot $\mathcal{A}_{a,b}(x,t|x_0)$ in Fig. \ref{fig:abs_prob}.

 \renewcommand{\theequation}{D\arabic{equation}}    
  \setcounter{equation}{0} 
\section*{Appendix D}
\label{sec:appD}
The expression for the survival probability studied in Sec. \ref{sec:split} requires knowledge of certain finite integrals. They are
\bea
\int_{a}^{b}{\rm d}x\,e^{-\beta |x-\alpha|}&=\mbox{sign}(b-\alpha)\frac{1-e^{-\beta |b-\alpha|}}{\beta}-\mbox{sign}(a-\alpha)\frac{1-e^{-\beta |a-\alpha|}}{\beta},
\label{eqn:int1}
\eea
and
\bea
 &\fl \hspace{70pt}\int_{a}^{b}{\rm d}x\,e^{-\delta |x-x_0|-\beta
 |x|}=e^{-\delta x_0}\int_{a}^{x_0}{\rm d}x\,e^{\delta x-\beta
 |x|}+e^{\delta x_0}\int_{x_0}^{b}{\rm d}x\,e^{-\delta x-\beta |x|} \nonumber \\
 &\fl =e^{-\delta x_0}\left[\frac{1-e^{\delta a-\beta |a|}}{\delta-\beta\, \mbox{sign}(a)}-\frac{1-e^{\delta x_0-\beta |x_0|}}{\delta-\beta\, \mbox{sign}(x_0)}\right]+e^{\delta x_0}\left[\frac{e^{-\delta x_0-\beta |x_0|}-1}{\delta+\beta\, \mbox{sign}(x_0)}+\frac{1-e^{-\delta b-\beta |b|}}{\delta+\beta\, \mbox{sign}(b)}\right]\nonumber \\
&\fl=e^{-\delta x_0}\left[\frac{1-e^{a[\delta -\beta\, \mbox{sign}(a)]}}{\delta-\beta\, \mbox{sign}(a)}-\frac{1-e^{ x_0[\delta -\beta\, \mbox{sign}(x_0)]}}{\delta-\beta\, \mbox{sign}(x_0)}\right]\nonumber \\
&\fl \hspace{150pt}+e^{\delta x_0}\left[\frac{e^{-x_0[\delta +\beta\, \mbox{sign}(x_0)]}-1}{\delta+\beta\, \mbox{sign}(x_0)}+\frac{1-e^{-b[\delta +\beta\, \mbox{sign}(b)]}}{\delta+\beta\, \mbox{sign}(b)}\right].
\label{eqn:int2}
\eea

For the exponential reset we have
\be
\fl \int_{a}^{b}{\rm d}x\,\widetilde{P}(x,\epsilon|a)=\frac{1}{\sqrt{2}(r+\epsilon)}\left\{\frac{r}{\epsilon}\left[2-e^{-b\sqrt{(r+\epsilon)/D}}-e^{a\sqrt{(r+\epsilon)/D}}\,\right]+1-e^{-(b-a)\sqrt{(r+\epsilon)/D}}\right\}
\label{eqn:int_prop_exp_reset}
\ee
and
\be
\fl \widetilde{\mathcal{G}}_{a}(b,\epsilon|x_0)=\frac{r\,e^{-b\sqrt{(r+\epsilon)/D}}+\epsilon\,e^{-(b-x_0)\sqrt{(r+\epsilon)/D}}-\widetilde{\mathcal{F}}(a,\epsilon|x_0)\left[r\,e^{-b\sqrt{(r+\epsilon)/D}}+\epsilon\,e^{-(b-a)\sqrt{(r+\epsilon)/D}}\,\right]}{r\,e^{-b\sqrt{(r+\epsilon)/D}}+\epsilon-\widetilde{\mathcal{F}}(a,\epsilon|b)\left[r\,e^{-b\sqrt{(r+\epsilon)/D}}+\epsilon\,e^{-(b-a)\sqrt{(r+\epsilon)/D}}\,\right]}.
\label{eqn:split_time_dep}
\ee
By taking the limit $\epsilon\rightarrow 0$ in Eq.
(\ref{eqn:split_time_dep}) it is straightforward to extract the
splitting probability shown in Eq. (\ref{eqn:split_prob_exp_reset})
since by construction $\widetilde{\mathcal{F}}(a,\epsilon\rightarrow
0|x_0)=\widetilde{\mathcal{F}}(a,\epsilon\rightarrow 0|b)=1$.
We have in fact that
\bea
\fl \widetilde{S}_{a,b}(\epsilon|x_0)=\frac{1}{v}
-\widetilde{\mathcal{G}}_a(b,\epsilon|x_0)\int_{a}^{b}{\rm
d}x\,\widetilde{P}(x,\epsilon|b)-\widetilde{\mathcal{G}}_b(a,\epsilon|x_0)\int_{a}^{b}{\rm
d}x\,\widetilde{P}(x,\epsilon|a),
\label{eqn:survival_2bound_exp}
\eea

\section*{Acknowledgements} We thank Nitant Kenkre for his continuous support and uncountable discussions on the paper. SG acknowledges the support and warm hospitality
of the University of Bristol. The work started as an activity under the Advanced
Study Group ``Statistical Physics and Anomalous Dynamics of Foraging" 
held in 2015 at the Max Planck Institute for the Physics of Complex Systems,
Dresden. This work was supported in part by the Engineering and Physical Sciences Research Council (EPSRC) UK Grant No. EP/I013717/1.

\section*{References}
\bibliographystyle{ieeetr}
\bibliography{Biblio_review}   

\end{document}